\documentclass[twocolumn,final,aps,prl,showpacs,superscriptaddress,amsmath,amssymb,amsfonts,floatfix]{revtex4-1}

\usepackage{graphicx}
\usepackage{multirow}
\usepackage{epsfig}
\usepackage{url}
\usepackage{color}
\usepackage{ulem}

\usepackage[colorlinks,breaklinks,bookmarks=true,citecolor=blue,linkcolor=blue,urlcolor=blue]{hyperref}
\usepackage{color}
\usepackage{tabularx}
\usepackage{sidecap}
\usepackage{soul}
\usepackage{float}

\usepackage{orcidlink}

\newcommand{\ls}[1]{{\color{black}#1}}

\begin{document}

\author{Liang Si\,\orcidlink{0000-0003-4709-6882}}
\email{liang.si@ifp.tuwien.ac.at}
\affiliation{School of Physics, Northwest University, Xi'an 710127, China}
\affiliation{Institute of Solid State Physics, TU Wien, 1040 Vienna, Austria}

\author{Paul Worm\,\orcidlink{0000-0003-2575-5058}}
\affiliation{Institute of Solid State Physics, TU Wien, 1040 Vienna, Austria}

\author{Dachuan Chen}
\affiliation{CAS Key Laboratory of Magnetic Materials and Devices \& Zhejiang Province Key Laboratory of Magnetic Materials and Application Technology, Ningbo Institute of Materials Technology and Engineering, Chinese Academy of Sciences, Ningbo 315201, China}

\author{Karsten Held\,\orcidlink{0000-0001-5984-8549}}
\email{held@ifp.tuwien.ac.at}
\affiliation{Institute of Solid State Physics, TU Wien, 1040 Vienna, Austria}

\title{Topotactic-hydrogen forms chains in $AB$O$_2$ nickelate superconductors}

\begin{abstract}
Despite enormous experimental and theoretical efforts, obtaining generally accepted conclusions regarding the intrinsic magnetic and electronic properties of superconducting nickelates remains exceptionally challenging. Experiments show a significant degree of uncertainty, indicating hidden factors in the synthesized films, which call for further investigations. 
\ls{One of those ``hidden factors'' is the possibility of intercalating hydrogen during the chemical reduction process from Nd(La)NiO$_3$ to Nd(La)NiO$_2$ using CaH$_2$. While hydrogen has been detected in experimental samples, not much is known about its distribution through the crystal and its influence on the electronic environment.}
Here, we \ls{show the tendency} toward the formation of one-dimensional hydrogen chains in infinite-layers LaNiO$_2$ superconductors using density-functional theory \ls{(DFT) supplemented by dynamical mean-field theory (DMFT)}. The formation of such hydrogen chains induces a coexistence of different oxidation states of Ni and competing magnetic phases, \ls{and possibly explains the recently observed charge order states in nickelate superconductors.} Furthermore, it contributes to the difficulty of synthesizing homogeneous nickelates and determining their ground states. 
The smoking gun to detect excess hydrogen in nickelates are flat phonon modes, which are \ls{infrared active} and quite \ls{insensitive} to the exact arrangement of the H atoms.
\end{abstract}

\maketitle

\section{I.~Introduction}
Efforts to achieve higher and higher superconducting transition temperatures already last for more than a century, starting with the original discovery of superconductivity in mercury by Kammerlingh Onnes. For the first 75 years, research focused on what we now call conventional superconductors. While their microscopic mechanism is understood by now within the framework of the  Bardeen, Cooper, and Schrieffer (BCS) \cite{BCS1957} theory, no superconductor at that time exceeded a transition temperature of 30\,K. This changed with the major breakthrough by Bednortz and M{\"u}ller \cite{Bednorz1986}, who discovered superconductivity in a copper oxide compound. Still, to this day, members of those cuprates hold the record for the transition temperature ($T_C$) at ambient pressure. It should be noted, though, that recently superconductivity around room temperature has been observed in hydrides, albeit under enormous pressure \cite{Drozdov2015,drozdov2019superconductivity,Somayazulu2022}.
While hydrides fall into the BSC class of superconductors, cuprates are distinctly different, and BSC generally fails to describe their $T_C$ \cite{Savrasov1996,Boeri2008}. Moreover, the pairing mechanism remains highly controversial even 35 years after the initial discovery of cuprate superconductivity. 
Many different mechanisms that account for the pairing of electrons have been proposed,   including antiferromagnetic (AFM) spin  \cite{Monthoux1991,Scalapino12,RevModPhys.84.1383,Vilardi2020,Sordi2012,Gull2015,Kitatani2019}
and various other fluctuations such as charge \cite{Castellani1997},  current \cite{Varma1997} and resonating valence bonds \cite{Anderson87}, possibly connected to the
the vicinity of a quantum critical point \cite{Chubukov1994,Castellani1997,Varma1997}.

However, a general consensus for the mechanism of superconductivity in strongly correlated superconductors has not been reached. Nevertheless, several common features of essentially all cuprates have been established: (i) superconductivity arises in square planar CuO$_2$ planes. (ii) Cu is in a 3$d^{9-\delta}$ state, where $\delta$ is the number of doped holes in the compound. (iii) There is one orbital at the Fermi-surface, which is a Cu-3$d_{x^2-y^2}$ and O-2$p$ hybrid in the spirit of the Zhang-Rice singlet \cite{Zhang1988}. (iv) Superconductivity is observed in a range of roughly $5-25\%$ hole-doping in the effective 3$d_{x^2-y^2}$ orbital, with details depending on the compound. 

Following these simple ingredients already early simulations have predicted the possibility of nickelates superconductors \cite{Anisimov1999} and  heterostructures thereof \cite{Chaloupka2008,PhysRevLett.103.016401,Hansmann2010b}. However, a superconducting nickelate was experimentally only confirmed in 2019 for Sr-doped NdNiO$_2$ by Li \emph{et al.} \cite{li2019superconductivity},  opening the door to a new playground of superconductivity after a long search. Nickelate superconductors have attracted intensive attention in the last three years, marked by an enormous theoretical and experimental activity, including, but not restricted to, Refs.~\cite{li2019superconductivity,zeng2020,Osada2020,Zeng2021,Osada2021,pan2021,PhysRevLett.125.077003,Motoaki2019,hu2019twoband,Wu2019,Nomura2019,Zhang2019,Jiang2019,Werner2019,Lechermann2019,Si2019,Lechermann2020,Petocchi2020,Adhikary2020,Subhadeep2020,Karp2020,Kitatani2020,Geisler2021,Klett2022,LaBolitta2022} on both infinite- and finite-layer nickelates \cite{pan2021,Worm2021c}.

Compared to cuprates, which are considered charge-transfer insulators with a Coulomb interaction larger than the charge-transfer gap ($U > \Delta$), the situation is reversed in nickelates, placing them instead in the Mott-Hubbard regime. Additionally, the Ni-O hybridization is reduced and the doped holes thus reside predominantly in the Ni $d$-band, specifically the $d_{x^2-y^2}$ orbital \cite{PhysRevB.104.L220505,krieger2021charge,goodge2021doping,Jiang2019}, while in cuprates they are \ls{predominately} located at the O sites. An additional aspect to consider is the Ni-La(Nd) hybridization \cite{Lee2004} and, in particular, the electron pockets at the $\Gamma$- and $A$-points. However,  they seem not to be essential for superconductivity. This was concluded based on calculations using the dynamical vertex approximation (D$\Gamma$A) \cite{Toschi2007,Held2008,Katanin2009,RMPVertex} for nickelates  \cite{Kitatani2020}, which is unbiased with respect to charge and spin fluctuations. These calculations find that spin fluctuations dominate and successfully predicted the superconducting dome prior to experiments in Nd$_{1-x}$Sr$_x$NiO$_2$ \cite{Li2020,zeng2020}. The recent measurements on defect-free (Nd,Sr)NiO$_2$ samples yield an even better agreement regarding $T_C$ and the dome structure as predicted by  D$\Gamma$A  \cite{lee2022character}.

While the parent compound of cuprates orders antiferromagnetically, there is still some uncertainty concerning the magnetic and electronic structure of nickelates. Resonant inelastic x-ray scattering (RIXS) \cite{krieger2021charge,tam2021charge} indicates the existence of charge density wave (CDW) order, including both the Ni and Nd sites. Moreover, the coexistence of Ni$^{1+}$ and Ni$^{2+}$ was reported; the latter is, however, incompatible  with monovalent Ni$^{1+}$ in LaNiO$_2$. 
Additionally, no long-range magnetic order  has been observed in LaNiO$_2$ and NdNiO$_2$, even in their bulk states \cite{lin2022universal,PhysRevResearch.4.023093}. Let us note at this point that while the Nd pockets seem not to be relevant for superconductivity, they quite distinctly alter the parent compound of nickelates compared to cuprates. Specifically, the pocket ``self-dopes'' the Ni 3$d_{x^2-y^2}$ orbital such that its filling is only $\sim$0.94 in the parent compound instead of $1$ as for cuprates. Hence, a comparison to $\sim$6\% hole-doped cuprates is more qualified. Indeed, such doping is already at the very edge, or even outside the AFM dome \cite{Keimer2015} in cuprates. Furthermore, magnetic excitations consistent with the AFM magnon dispersion, which is expected for doped cuprate-like AFM order, have recently been observed in NdNiO$_2$ films by RIXS \cite{Lu2021}. Similarly, nuclear magnetic resonance (NMR) indicates AFM fluctuations \cite{Cui2021}.
The $\mu$SR and low-field static and dynamic magnetic susceptibility measurements performed by Ortiz \emph{et al.} reveal the presence of short-range magnetic correlations and glassy spin dynamics that hint at a possible (local) magnetic-frustration \cite{PhysRevResearch.4.023093}. These effects have been attributed to local oxygen (O) nonstoichiometry. \ls{Another  $\mu$SR measurement \cite{fowlie2022intrinsic} shows that AFM correlations or a magnetically long-ranged ordered ground state persist within the superconducting phase.}

The above work reveals (1) the coexistence of various oxidation states and (2) various magnetic signals, which were recently observed to be (short-range) AFM correlations. These measurements strongly highlight further similarities and differences between nickelates and cuprates. Previous theoretical  \cite{Si2019,si2022fingerprints,Malyi2022,Puphal2022,PhysRevMaterials.6.044807} and experimental \cite{Cui2021} studies demonstrated the possibility of intercalating topotactic-hydrogen (H) defects and their consequences in nickelates. Those theoretical studies focused primarily on the hydrogen saturated compound, i.e. (Nd)LaNiO$_2$H. A natural question arising is how \ls{a more realisitc, lower density H defects} will arrange itself in the crystal and alter the magnetic and electronic properties of the corresponding synthesized crystals. Especially interesting is here the question of how much of the observed experimental uncertainties can be explained and attributed to such defects.

In this paper, we go beyond \cite{Si2019} that reported $E_b$ of various $AB$O$_2$ compounds by investigating $E_b$ of LaNiO$_2$H$_{\delta}$ systems with $\delta$=0\%-100\%. The structural, electronic and magnetic properties of LaNiO$_2$H$_{\delta}$ are studied using density-functional theory (DFT)  \cite{PhysRev.136.B864} and dynamical mean-field theory (DMFT) calculations \cite{kotliar2004strongly,PhysRevLett.62.324,RevModPhys.68.13,held2007electronic}. 
A special behavior of topotactic-H is evidenced by our computations: topotactic-H in infinite layers tends to form one-dimensional (1D) chains along the $z$-direction. The captured H atoms are mainly confined by the Ni-sublattice while the La(Nd)-sublattice plays the role of distributing the H-chains along the (110)-direction by forming La-$d_{xy}$-H-1$s$ bonds. Such 1D H-chains affect the local properties by leading to various oxidation states in Ni and enhancing the 3D character of magnetism, giving rise to the inhomogeneity and uncertainty of experimental observations. \ls{Finally, we predict that the formation and existence of 1D H-chains can be detected by techniques such as RIXS and infrared spectroscopy, with the emergence of flat phonon modes being an important indicator.}

The paper is organized as follows:  Section~II provides the computational details and crystal structures considered; in Section~III we present corresponding results and discussions: Specifically, Sections~III~A, B, C, D, E, and F are dedicated to the results and discussions of (A) structural, (B) energetic, (C) dynamical (phonon), (D) electronic, (E) Fermi-surface and (F) magnetic properties, respectively.  Section~IV presents a conclusion and outlook.

\section{II.~Methods}
\label{sec:methods}
\begin{figure}[t]
\centering
\includegraphics[width=0.5\textwidth]{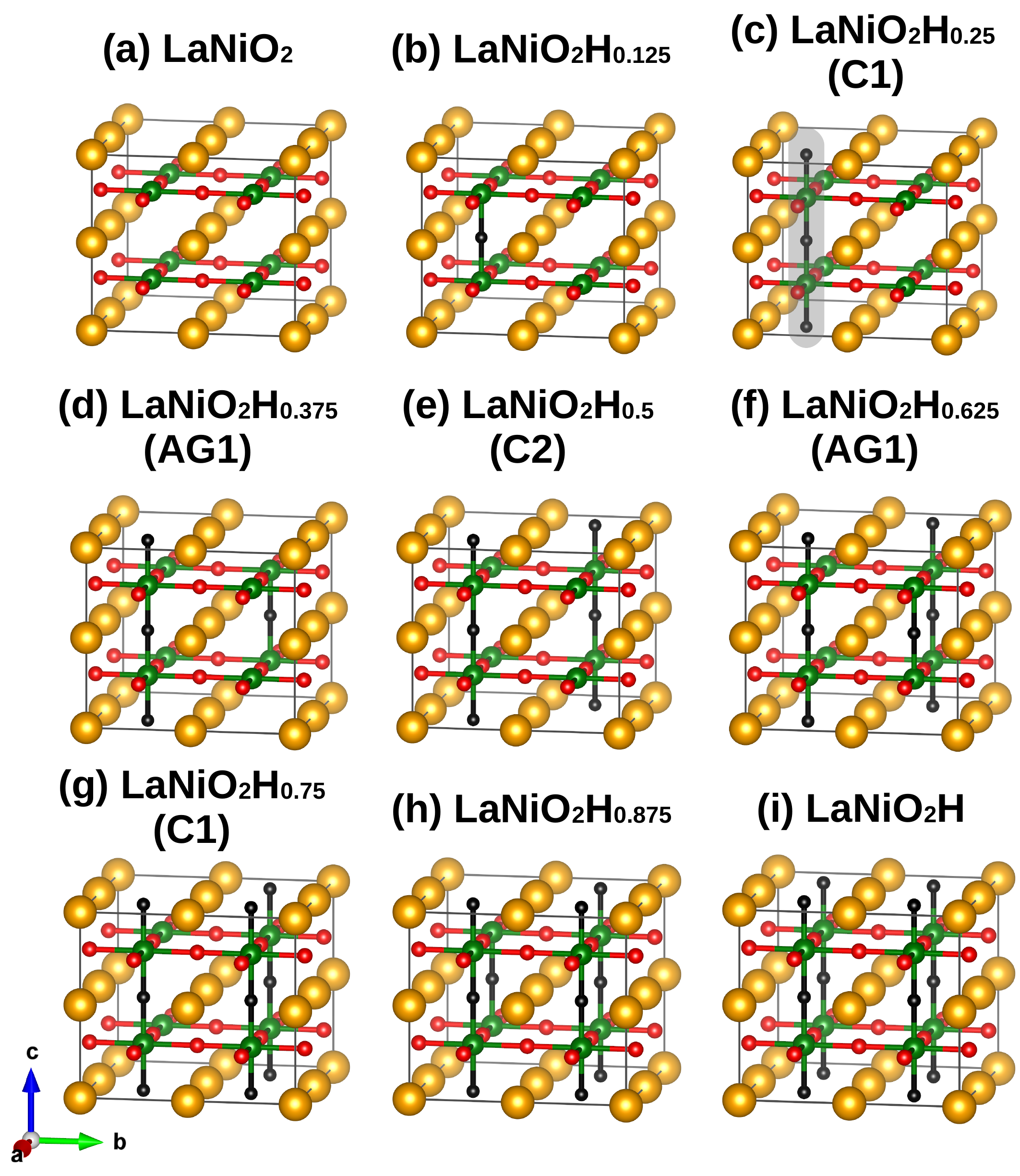}
\caption{Ground state crystal structures of LaNiO$_2$H$_{\delta}$ ($\delta$=0.0\%-100\%; 
yellow atoms: La; red: O; green: Ni; black: H).
All other possible structures for each topotactic-H concentration are shown in Appendix~I Fig.~\ref{FigA2-1}, \ref{FigA2-2} and \ref{Fig7_Structure}. The compounds without H and with fully topotactic-H correspond to (a) LaNiO$_2$ and (i) LaNiO$_2$H. For (c) LaNiO$_2$H$_{0.25}$, the grey region highlights the H-chain. The terms in brackets are the names of the corresponding structures as in Fig.~\ref{Fig7_Structure}.}
\label{Fig1}
\end{figure}

The ground state crystal structures with the lowest total energy of LaNiO$_2$H$_{\delta}$ ($\delta$=0.0\% to 100\%) are shown in Fig.~\ref{Fig1}. To obtain these results, DFT-level structural relaxations and static calculations 
are performed using the \textsc{Vasp} code \cite{PhysRevB.47.558,kresse1996efficiency} with the Perdew-Burke-Ernzerhof version for solids of the generalized gradient approximation (GGA-PBESol) \cite{PhysRevLett.100.136406} and a dense 13$\times$13$\times$15 $k$-mesh for the unit cells of $\delta$=0 and 100\% (LaNiO$_2$ and LaNiO$_2$H), and a 9$\times$9$\times$11 $k$-mesh for the 2$\times$2$\times$2 supercell of $\delta$=0\%, 12.5\%, 25\%, 37.5\%, 50\%, 62.5\%, 75\%, 87.5\%, 100\%. \ls{For each H concentration, structural relaxation and static total energy calculations for all possible inequivalent combinations of apical hydrogen distribution in a 2$\times$2$\times$2 supercell 
\footnote{According to our previous work \cite{Si2019} and preliminary DFT calculations as shown in Appendix~I, the out-of-plane topotactic-H is energetically more favorable compared with the in-plane topotactic-H. Hence we limit all the possible topotactic-H positions at out-of-plane positions (apical position above/under Ni ions), i.e., the positions created by removing O from LaO layers.} 
are carried out and shown in Fig.~\ref{Fig1} (and Fig.~\ref{Fig7_Structure}-\ref{FigA2-2} and Table~\ref{Tab:energy} in Appendix~I).} Phonon spectra calculations are performed with the frozen phonon (finite displacement) method using the \textsc{Phonopy} code \cite{togo2015first} interfaced with \textsc{Vasp} for the relaxed ground state structures of each H concentration.

In both our previous theoretical study \cite{Si2019} and in this article, the binding energy $E_b$ of hydrogen atoms is computed as:
\begin{equation}
E_{b}=\{E[{\rm LaNiO}_2] + n \times \mu[{\rm H}] - E[{\rm LaNiO}_2{\rm H}_{\delta}]\}/n.
\label{Eq1}
\end{equation}
Here, $E[{\rm LaNiO}_2]$ and $E[{\rm LaNiO}_2{\rm H}_\delta]$ are the total energy of the infinite-layer LaNiO$_2$ and the hydride-oxides LaNiO$_2$H$_\delta$ supercell consisting of 2$\times$2$\times$2 chemical units \ls{(for tests employing larger supercells see Fig.~\ref{FigA1} in Appendix~I)}, while $\mu[H]=E[H_2]/2$ is the chemical potential of H, and $n=\delta \times 2^3$ is the number of H atoms in the supercell. A positive (negative) $E_b$ indicates that the topotactic H process is energetically favorable (unfavorable) and LaNiO$_2$H$_\delta$ will be formed instead of LaNiO$_2$ and H$_2$/2.

DFT-level electronic and magnetic structures are subsequently investigated using \textsc{Wien2K} \cite{blaha2001wien2k,Schwarz2003} with the PBE \cite{PhysRevLett.77.3865} version of GGA on a dense $k$-mesh with a total of 3000 $k$-points. As an input for the DMFT calculations, a low-energy effective Hamiltonian is generated by projecting the La-5$d$+Ni-3$d$ derived DFT bands computed by \textsc{WIEN2K} around the Fermi level onto Wannier functions \cite{PhysRev.52.191,RevModPhys.84.1419} using {WIEN2WANNIER}  \cite{mostofi2008wannier90,kunevs2010wien2wannier}. These are supplemented by a local Kanamori interaction and we employ the fully localized limit as double counting correction scheme \cite{Anisimov1991}. Constrained random phase approximation (cRPA) \cite{PhysRevB.77.085122} calculations motivate
an intraorbital Hubbard interaction $U$=4.4\,eV and a Hund's exchange $J$=0.65\,eV (inter orbital $U'$=$U$-2$J$=3.1\,eV) for Ni-3$d$, $U$=2.5\,eV and $J$=0.25\,eV ($U'$=2.0\,eV) for La-5$d$ orbitals.
\ls{The interaction parameters obtained using cRPA for both the Ni-3$d$ and La-5$d$ orbitals are consistent with relevant research prior to this study including, but not limited to Refs.~\cite{PhysRevResearch.2.033445,PhysRevLett.125.077003,PhysRevLett.101.076402,PhysRevB.103.165133,PhysRevLett.125.166402,PhysRevB.90.165105}.}
We solve the resulting many-body Hamiltonian at room temperature (300\,K) within DMFT employing a continuous-time quantum Monte Carlo solver in the hybridization expansions \cite{RevModPhys.83.349} using \textsc{W2dynamics} \cite{PhysRevB.86.155158,w2dynamics2018}. Real-frequency spectra are obtained with the \textsc{ana\_cont} code \cite{Kaufmann2021} via analytic continuation using the maximum entropy method (MaxEnt) \cite{PhysRevB.44.6011,PhysRevB.57.10287}.

\section{III. Results and Discussion}
\label{sec:results}
\subsection{A.~Formation of one-dimensional H-chains}
\label{sec:formation}
In Fig.~\ref{Fig1} we show the ground state crystal structure with the lowest total energy for each concentration of topotactic-H LaNiO$_2$H$_{\delta}$ ($\delta$=0.0\% to 100\%). The case with 0\% and 100\% topotactic-H corresponds to LaNiO$_2$ [Fig.~\ref{Fig1}(a)] and LaNiO$_2$H [Fig.~\ref{Fig1}(i)], respectively. 
The simulation of the process of gradually increasing the concentration of topotactic-H in LaNiO$_2$ is achieved by intercalating H atom(s) to the pristine 2$\times$2$\times$2 LaNiO$_2$ supercell in Fig.~\ref{Fig1}(a). This process enables us to investigate the energetic, crystal, electronic and magnetic properties for LaNiO$_2$H$_{\delta}$ with $\delta$=0.0\%, 12.5\%, 25.0\%, 37.5\%, 50.0\%, 62.5\%, 75.0\%, 87.5\% and finally 100\%. For each concentration, the ground state crystal structure is predicted by computing the total DFT energy of all possible structures \ls{where H is intercalated topotactically at perovskite oxygen positions}. For the details of possible structures and $E_b$ see Appendix~I Fig.~\ref{Fig7_Structure}-\ref{FigA2-2} and Table~\ref{Tab:energy}.

For the compounds LaNiO$_2$ [Fig.~\ref{Fig1}(a)], LaNiO$_2$H$_{0.125}$ [Fig.~\ref{Fig1}(b)], LaNiO$_2$H$_{0.875}$ and LaNiO$_2$H, the ground state structures are not named as there is only one inequivalent structure for these cases in the 2$\times$2$\times$2 supercell. In Fig.~\ref{Fig1}(b,c), the formation of 1D H-chains can clearly be seen: the second H atom [Fig.~\ref{Fig1}(c)] is energetically favorable to occupy the apical position on top of the first H atom [Fig.~\ref{Fig1}(b)], thus forming a 1D H-chain. Similar to Fig.~\ref{Fig1}(b,c), Fig.~\ref{Fig1}(d,e) shows the process of the formation of the second H-chain when the third and fourth H atoms are intercalated into the supercell. One notable observation is the favorable position for the third and fourth H atom. Both the single third H atom [Fig.~\ref{Fig1}(d)] and the second H-chain [fourth H atom; Fig.~\ref{Fig1}(e)] are energetically favorable to occupy the empty space in the (110) direction instead of (100)/(010). Combining this tendency with the formation of 1D H-chain in Fig.~\ref{Fig1}(b,c), we conclude that the formation of the 1D H-chains is driven by the bonds between H-1$s$ and Ni-$d_{z^2}$ orbitals, and the bonds between H-1$s$ and La-$d_{xy}$ play important roles concerning the detailed arrangement of the H-chains. Finally, further chains are formed when intercalating more H in Fig.~\ref{Fig1}(f-i).

\subsection{B.~Topotactic-H reshapes the magnetic structures}
\label{sec:topotactic}
\begin{figure*}[tb]
\includegraphics[width=0.9\textwidth]{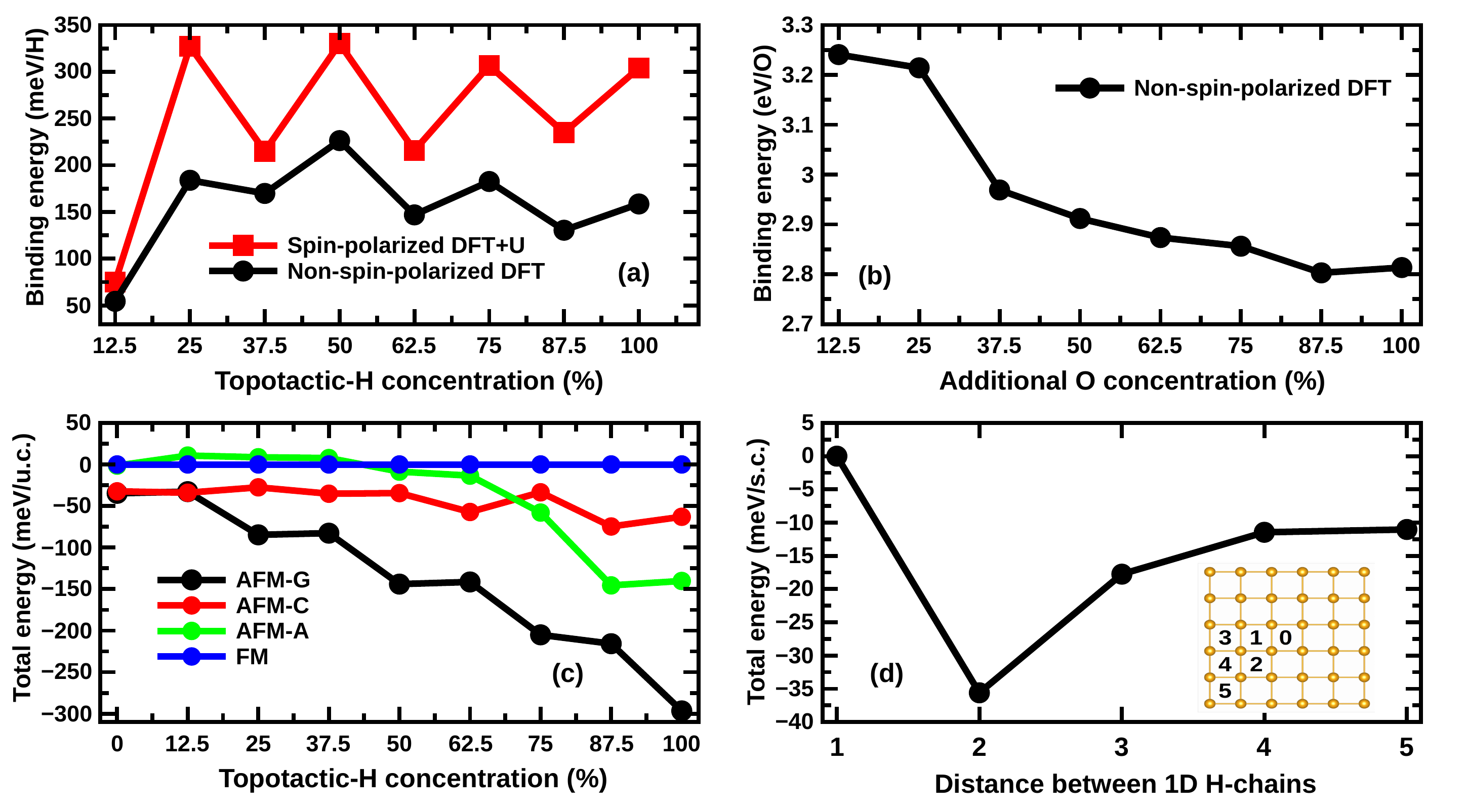}
\caption{DFT and DFT+$U$ results for LaNiO$_2$H$_{\delta}$. (a) H binding energy ($E_b$) of LaNiO$_2$H$_{\delta}$ from non-spin polarized DFT and spin-polarized DFT+$U$ calculations. (b) Non-spin polarized DFT calculations of oxygen (O) binding energy for LaNiO$_{2+\delta}$. (c) DFT+$U$ magnetically total energies of LaNiO$_2$H$_{\delta}$ for four different magnetic states: ferromagnetic (FM), $A$-, $C$- and $G$-type antiferromagnetic orders (AFM-$A$, AFM-$C$ and AFM-$G$). The total energy of FM state is set to zero for comparison. (d) DFT total energy calculations for a 5$\times$5$\times$1 LaNiO$_2$ supercell with two 1D H-chains at different positions: at ``0'' between four yellow La atoms in the top view, we locate the first H-chain, and ``1-5'' indicates the first, second, third, fourth and fifth nearest positions for the second H-chain. The energy of the nearest neighbor arrangement ``1" is set to zero.}
\label{Fig2}
\end{figure*}

To support the hypothesis concerning the formation of 1D H-chains in nickelate superconductors, we perform DFT-level calculations on the topotactic-H binding energy $E_b$ for all ground state structures (as shown in Fig.~\ref{Fig1}). Fig.~\ref{Fig2}(a) shows $E_b$ of different topotactic-H concentrations. Beyond Ref.~\cite{Si2019} that reports $E_b$ obtained from non-spin-polarized DFT calculations without a DFT+$U$ treatment, here we also performed spin-polarized DFT+$U$ calculations, and the magnetically spin-polarized total energies are used to compute $E_b$ according to Eq.~(\ref{Eq1}). Such setups are closer to the real conditions within nickelates, where local Ni moments form.

As shown in Fig.~\ref{Fig2}(a), $E_b$ reaches a local maximum when the topotactic-H concentration is 25\%, 50\%, 75\% and 100\%, corresponding to 1, 2, 3 and 4 H-chains in the supercell. Please note, according to Eq.~(\ref{Eq1}), the local maximum of $E_b$ indicates the energetically most stable structures of LaNiO$_2$H$_{\delta}$.  Hence the results in Fig.~\ref{Fig2}(a) demonstrate that topotactic-H atoms in infinite-layer nickelates tend to form 1D H-chains, especially \ls{when $E_b$ is} obtained from spin-polarized DFT+$U$ computations. In this context, another important question is: when there is a certain amount of H atoms remaining in samples, will they form \ls{additional} H-chains (e.g.~in the structure of LaNiO$_2$H$_{0.5}$) or a locally full topotactic phase (i.e.~LaNiO$_2$H)? For non-spin-polarized (DFT) and spin-polarized (DFT+$U$) calculations, $E_b$ of LaNiO$_2$H$_{0.5}$ is always higher than that of LaNiO$_2$H, by 68\,meV and 27\,meV per H, respectively, which corresponds to temperatures of 789\,K and 313\,K. This consistently indicates that the formation of H-chains is energetically favorable even at room temperature, although at such elevated temperatures, some degree of disorder can be expected.

To show that this 1D H-chain formation is unique, i.e. other intercalated elements fail to form such chains, we also compute the binding energy $E_b$ for O. In fact, \ls{one can easily imagine that excess oxygen remains in the film after an (insufficient)} reduction process from LaNiO$_3$ to LaNiO$_2$ under the participation of CaH$_2$. The energy cost in the process from LaNiO$_3$ (i.e., $\delta$=100\% in LaNiO$_{2+\delta}$) to LaNiO$_2$ ($\delta$=0\%) reflects the difficulty of removing O atoms gradually as the binding energy increases with decreasing O concentration. However, this trend is monotonous as shown in Fig.~\ref{Fig2}(b). That is, the $E_b$ of O is only sensitive to the density of O in LaNiO$_{2+\delta}$. This is distinctly different from the topotactic-H case (LaNiO$_2$H$_{\delta}$). \ls{We perform additional DFT calculations employing a larger supercell from 3$\times$3$\times$3 to 5$\times$5$\times$5, the results are shown in Appendix~I and Fig.~\ref{FigA1}. We find that the $E_b$ of infinite H-chains in 3$\times$3$\times$3 is maximal for all tested configurations. The corresponding structural and chemistry periodic vector $\vec{R}$=[$\frac{1}{3}$, $\frac{1}{3}$, 0] is consistent with the recently observed charge order state in nickelate superconductors \cite{krieger2021charge,tam2021charge}.}

As discussed in Ref.~\cite{Si2019}, a complete intercalation of H triggers a transition from a quasi-2D strongly correlated single-band ($d_{x^2-y^2}$) metal to a two-band ($d_{x^2-y^2}$+$d_{z^2}$) 3D AFM Mott-insulator. Here, by changing the density of topotactic-H we investigate how the formation of 1D H-chains reshapes the magnetic order of LaNiO$_2$H$_{\delta}$. Fig.~\ref{Fig2}(c) shows the spin-polarized total energy per unit cell of LaNiO$_2$H$_{\delta}$ obtained from DFT+$U$ calculations by setting the supercell in Fig.~\ref{Fig1} to ferromagnetic (FM), $A$-, $C$- and $G$-type AFM orders. For LaNiO$_2$ ($\delta$=0\%), the AFM-$C$ and AFM-$G$ state have similar total energies, which are $\sim$32\,meV/Ni lower than those of FM and AFM-$A$, indicating that the system is dominated by its quasi-2D character with strong intra-layer coupling, but only weak inter-layer coupling ($\sim$2\,meV/Ni). At a doping of $\delta$=12.5\%, the AFM-$C$ state becomes the magnetic ground state being $\sim$1.3\,meV/Ni lower in energy than AFM-$G$. However, such a small energy difference is likely not strong enough to pin the AFM-$C$ state and can thus lead to magnetic frustration, including suppression of long-range order, which could lead to a spin-ice or glassy behavior.
As $\delta$ further increases, AFM inter-layers coupling starts to become dominant and for $\delta>$12.5\% AFM-$G$ becomes energetically favorable. Between $\delta$=12.5\%-100\%, the energy curve of AFM-$G$ shows a step-like tendency, with remarkable energy gains at $\delta$=25\%, 50\%, 75\% and 100\%, indicating that the formation of the 1D H-chains causes this energy gain. Another notable change with H concentration occurs at $\delta$$>$$75$\%, where AFM-$A$ order starts to be favorable compared to AFM-$C$ by $\sim$77\,meV/Ni, indicating that the inter-layer coupling increases in strength as the topotactic-H concentration increases. This observation is in agreement with the observed increase of three-dimensionality as we go toward the fully intercalated compound. 

We further investigate how multiple H-chains will arrange themselves in infinite-layer nickelates. This simulation is done by employing a 5$\times$5$\times$1 supercell and locating 2 H-chains at different positions. As shown in the inset of Fig.~\ref{Fig2}(d), the second H-chain tends to occupy the next site along the (110) crystal cell vector (``2") rather than along the (100)/(010) direction (``1"). In fact, the nearest-neighbor location is the place where the second chain wants to be the least, evidenced by a relative energy gain of $\sim 11$\,meV for locations further away (``5").
This result reflects the importance of bonds between La-$d_{xy}$ and H-1$s$ regarding the distribution of H-chains when the local density of topotactic-H does not reach 100\%. This conclusion will be further discussed in the context of phonon calculations, see Fig.~\ref{Fig3} below. \ls{As shown  in Fig.~\ref{FigA1} of Appendix~I, energetically even more favorable is the arrangement of the H-chains at a distance of three unit cells, which is at odds with a periodic arrangement in a 5$\times$5$\times$1 supercell.}


\subsection{C.~Phonon vibrations induced by topotactic-H}

\begin{figure*}[tb!]
\includegraphics[width=0.85\textwidth]{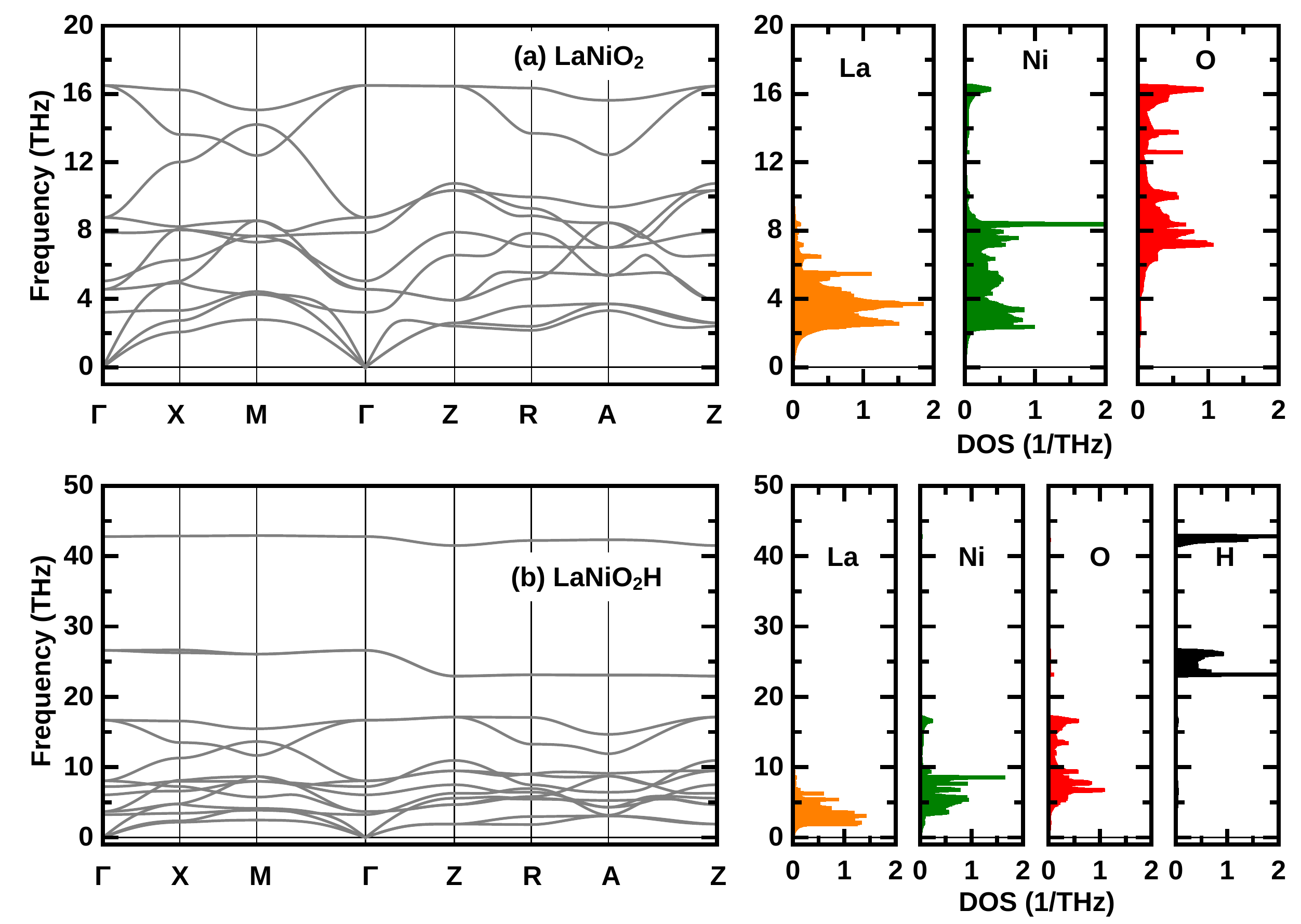}
\caption{DFT phonon spectra of (a) LaNiO$_2$ and (b) LaNiO$_2$H. Left panels: phonon dispersion reproduced from \cite{si2022fingerprints}. Right panels:  site-resolved phonon density of states. Please note the different scales of the $y$-axis in (a) and (b). }
\label{Fig3}
\end{figure*}

To further investigate the effects induced by topotactic-H, we perform phonon calculations with the frozen phonon method (finite displacement method) using the \textsc{Phonopy} \cite{togo2015first} code interfaced with \textsc{Vasp} and recapitulate the results of Ref.~\cite{si2022fingerprints} for  completing the physical picture.
As shown in Fig.~\ref{Fig3}(a), LaNiO$_2$ is dynamically stable, being consistent with the previous report on NdNiO$_2$ \cite{Nomura2019}. Its highest frequency optical phonon mode at around 14 to 16\,THz has also been observed in recent experimental resonant inelastic x-ray scattering (RIXS) measurements \cite{krieger2021charge} showing a weakly dispersive optical phonon at $\sim$60\,meV ($\sim$15\,THz).

From the right panels of Fig.~\ref{Fig3}(a), one can see the low-frequency region (0-6\,THz) is mainly stemming from La, while Ni contributes to the signal at 2-8\,THz and the high-frequency region (6-17\,THz) is from O atoms. This can easily be understood as La is the heaviest atom and O is the lightest one among the three. For the phonon spectra of LaNiO$_2$H$_{\delta}$ at other H concentrations, see Appendix~II and Fig.~\ref{Fig8_phonon}.

Fig.~\ref{Fig3}(b) shows, in comparison, the phonon mode of LaNiO$_2$H ($\delta$=100\% for LaNiO$_2$H$_{\delta}$), which is also dynamically stable as evidenced by its real frequencies. Please note that the phonon dispersions between 0 to 20\,THz are essentially the same as those in LaNiO$_2$, indicating that the phonon modes of topotactic-H and the LaNiO$_2$-superlattice are well decoupled. As shown in the phonon density of states (right panel), topotactic-H only contributes to two distinct high-frequency modes at $\sim$25 and 43\,THz, corresponding to 103 and 178\,meV, respectively. From our phonon eigenvector analysis, the single peak at 43\,THz is formed by the vibration of H atoms along (001)-direction ($z$-direction). In comparison, the double peaks at 25\,THz are generated by the double degenerate in-plane (100)/(110) vibrations ($xy$-plane). The formation of the double degenerate modes around $\sim$25\,THz is because there is always another symmetrically protected orthogonal direction for the vibrations along the $xy$-plane.

We explain these phonon modes in detail by computing the bonding strength between H-1$s$-Ni-$d_{z^2}$ and H-1$s$-La-$d_{xy}$. Our tight-binding calculations yield an electron hopping term of -1.604\,eV between H-1$s$ and Ni-$d_{z^2}$, while it is only -1.052\,eV from La-$d_{xy}$ to H-1$s$. That is, the larger H-1$s$-Ni-$d_{z^2}$ overlap leads to a stronger $\delta$-type bonding and, together with the shorter $c$-lattice constant, to a higher phonon frequency. Additionally, the shorter $c$-lattice in LaNiO$_2$ is also expected to increase the strength of the H-1$s$-Ni-$d_{z^2}$ bond.

\ls{The modes at high energies, i.e., 43 and 25\,THz, describe vibrations of the light hydrogen atoms out-of-plane and in-plane, respectively. Since hydrogen is essentially H$^{-}$ here, this movement is connected with a change of dipole moment. For this reason the modes couple directly to light, they are infrared active. On the other hand, the polarizability is symmetric with respect to a positive or negative movement of the H atoms out of the equilibrium position, hence they are Raman inactive.}

So far, we have obtained several important conclusions concerning properties and arrangements of topotactic-H in infinite-layers nickelate superconductors: (1) our previous band character computations for LaNiO$_2$H \cite{Si2019} report that the H-1$s$ band at $\sim$-7\,eV (-2\,eV) is composed of an H-1$s$ and Ni-$d_{z^2}$ bonding (anti-bonding) state; (2) the Ni-$d_{z^2}$-H-1$s$ bond plays the most important role regarding the formation of 1D H-chains; (3) the La-$d_{xy}$-H-1$s$ bond is decisive for the structural arrangement of \ls{different} H-chains; (4) the Ni-sublattice is more relevant for capturing topotactic-H atoms than the La-sublattice; \ls{(5) the phonon modes are possible indicators for detecting topotactic-H in infrared spectroscopy or RIXS measurements. This calls for further going optical spectroscopy with thicker nickelate films and bulk crystals, beyond \cite{cervasio2022optical} where the spectrum is still substrate dominated.}

\subsection{D.~Electronic structure of LaNiO$_2$H$_{\delta}$}

\begin{figure}[h]
\includegraphics[width=1.0\columnwidth]{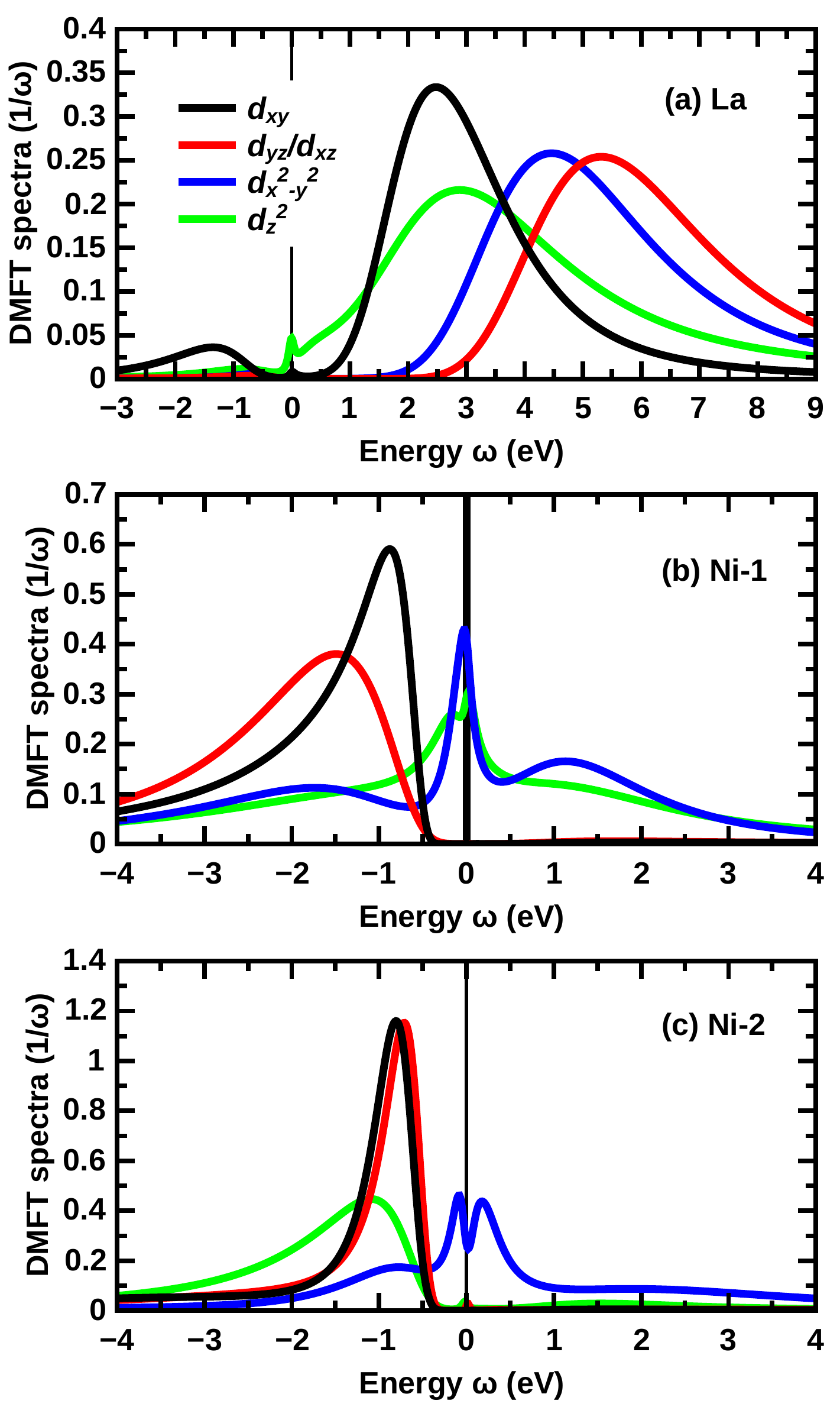}
\caption{DMFT $k$-integrated spectral functions $A$($\omega$) of LaNiO$_2$H$_{0.5}$ (in units of 1/$\omega$), resolved for (a) La, (b) first Ni site (Ni-1) and (c) second Ni site (Ni-2).}
\label{Fig4_H4_2}
\end{figure}

In this subsection, we study the electronic structure of LaNiO$_2$H$_{\delta}$ ($\delta$=0.5).  Fig.~\ref{Fig4_H4_2} shows the $k$-integrated spectral functions $A$($\omega$) of LaNiO$_2$H$_{0.5}$ obtained from DMFT calculations. 
Note that in this setup, there are only two inequivalent Ni sites and one Nd site (see Fig.~\ref{Fig5_H4_FS_DMFT} for structural details).
In this respect, we define the ``normal'' Ni, which is in the center of the NiO$_4$ square, as ``Ni-2'' and the ones that are between topotactic-H as ``Ni-1''. As shown, the La-Ni hybridization is not fully eliminated by H-chains. This is  in contrast to the fully H-intercalated LaNiO$_2$H (see Ref.~\cite{Si2019}). In particular, the  La density of states  below the Fermi energy in Fig.~\ref{Fig4_H4_2}(b) originates mainly from the La-$d_{xy}$ hybridization in the energy range  -3 to -1\,eV and  a La-$d_{z^2}$ pocket at the Fermi energy (0\,eV\ls{; cf.~Fig.~\ref{Fig5_H4_FS_DMFT}}). The former is a consequence of the $A$-pocket \cite{PhysRevLett.125.077003}, which is mainly composed by La-$d_{xy}$, and the latter is from the hybridization between Ni-$d_{z^2}$ and La-$d_{z^2}$ at the $\Gamma$-momentum. The hybridization also reflects in a larger  occupation per La which is 0.24 electrons/site, see Table~\ref{Tab:occupation}.

As shown in Fig.~\ref{Fig4_H4_2}(b) and (c), Ni-2 hosts a single-band $d_{x^2-y^2}$ picture while Ni-1 has to be described by a two-band model including both $d_{x^2-y^2}$ and $d_{z^2}$. Unlike the two-band Mott-insulating state in LaNiO$_2$H \cite{Si2019}, Ni-1 is a two-band correlated metal with a strongly renormalized effective mass $m^*$/$m_b$ (see Table~\ref{Tab:occupation}). As Ni-2 has a strongly correlated $d_{x^2-y^2}$ orbital ($n$$\sim$0.83) and a (nearly) fully filled $d_{z^2}$ orbital ($n$ $\sim$1.96), its configuration is close to Ni$^{1+}$ and 3$d^9$ \footnote{Note that the occupation of the (mainly empty) La orbitals is enhanced by La-Ni hybridization since the predominately Ni orbitals below the Fermi energy admix some La character, for detailed discussion in this aspect see \cite{Held2022,Worm2021c}}.

\ls{As shown in Table~I for LaNiO$_2$H$_{0.5}$, topotactic-H absorbs electrons mainly from the neighboring Ni-1 ion, specifically its $d_{z^2}$-orbital whose occupation is essentially reduced from 2 to 1. The thus half-filled Ni-1 $d_{z^2}$-orbital and half-filled $d_{x^2-y^2}$-orbital form a spin-1  due to Hund's exchange, which also drives both orbitals towards a more equal occupation. Actually it increases the occupation of the $d_{x^2-y^2}$-orbital slightly, as a secondary effect. Another secondary effect is that with the hydrogen atom absorbing electrons, as a matter of course also the Ni-2 is affected, albeit much less so, by removing about 0.1 electrons and here from the $d_{x^2-y^2}$-orbital.}

As reported in previous research \cite{Si2019}, 100\% topotactic-H eliminates the hybridization between La(Nd)-$d$ and Ni-$d$ by reversing (shifting up) the $A$-pocket ($\Gamma$-pocket) in LaNiO$_2$H. Hence an open question is up to which doping the pockets and the La-$d$-Ni-$d$ hybridization survive in LaNiO$_2$H$_{\delta}$ (0$<$$\delta$$<$1). 
Table~\ref{Tab:occupation} shows the La-site occupation under different concentrations of topotactic-H.
With increasing the concentration of topotactic-H, the electron occupation at La, which is a consequence of La-$d$-Ni-$d$ hybridization and the La-derived pockets, is remarkably reduced. 

The above computations in fact explain the recent experimental observations of the existence of Ni$^{2+}$ in undoped LaNiO$_2$ \cite{tam2021charge,krieger2021charge}. Existing 1D H-chains reconstruct a 3D character of the Ni sites between topotactic-H without affecting other Ni sites. This allows for the coexistence of Ni$^{2+}$ and Ni$^{1+}$. Unlike in the 100\% case (LaNiO$_2$H), 1D H-chains allow a slight La-5$d$-Ni-3$d$ hybridization, which has been extensively observed in recent resonant inelastic X-ray scattering (RIXS) measurements \cite{Hepting2020,tam2021charge,krieger2021charge,Lu2021}. Additionally, the existence of 1D H-chains results in a charge ordering state induced by disproportionate Ni oxidation states (nominal 3$d^9$ Ni$^{1+}$ and high-spin 3$d^8$ Ni$^{2+}$). Lastly, the uncertainty in the concentration of topotactic-H may contribute to the Hall coefficient \cite{li2019superconductivity,Li2020,zeng2020}, as it induces a \ls{mixture}  of single/two/multi bands that are composed by $d_{x^2-y^2}$/$d_{x^2-y^2}$+$d_{z^2}$/$d_{x^2-y^2}$+$d_{z^2}$+$d_{xy}$/$d_{z^2}$(of La) near $E_f$, respectively.

\begin{table}[h]
\caption{ DMFT occupation $n$ at (a) La site for different concentrations $\delta$ of topotactic-H and (b) effective mass enhancement ($m^*$/$m_b$) and occupation $n$ of different Ni orbitals for LaNiO$_2$H$_{0.5}$.}

\begin{tabular}{c|c|c|c|c}
\hline
\hline
(a) LaNiO$_2$H$_{\delta}$ \\
\hline
$\delta$ &0.25 & 0.50 & 0.75  & 1.00 \\
\hline
$n$(La-site) & 0.32 & 0.24 & 0.20 & 0.00  \\
\hline
\hline
\noalign{\smallskip}\noalign{\smallskip}
\noalign{\smallskip}\noalign{\smallskip}
\hline
\hline
(b) LaNiO$_2$H$_{0.5}$ \\
\hline
orbital & Ni-1 $d_{x^2-y^2}$  & Ni-1 $d_{z^2}$  & Ni-2 $d_{x^2-y^2}$  & Ni-2 $d_{z^2}$  \\
\hline
m$^*$/m$_b$ & 4.13 & 2.85 & 2.77 & 1.19 \\
$n$ & 1.05 & 1.11 & 0.83 & 1.96 \\
\hline
\hline
\end{tabular}
\label{Tab:occupation}
\end{table}

\subsection{E.~Fermi surface of LaNiO$_2$H$_{\delta}$}

\begin{figure*}[t]
\includegraphics[width=0.95\textwidth]{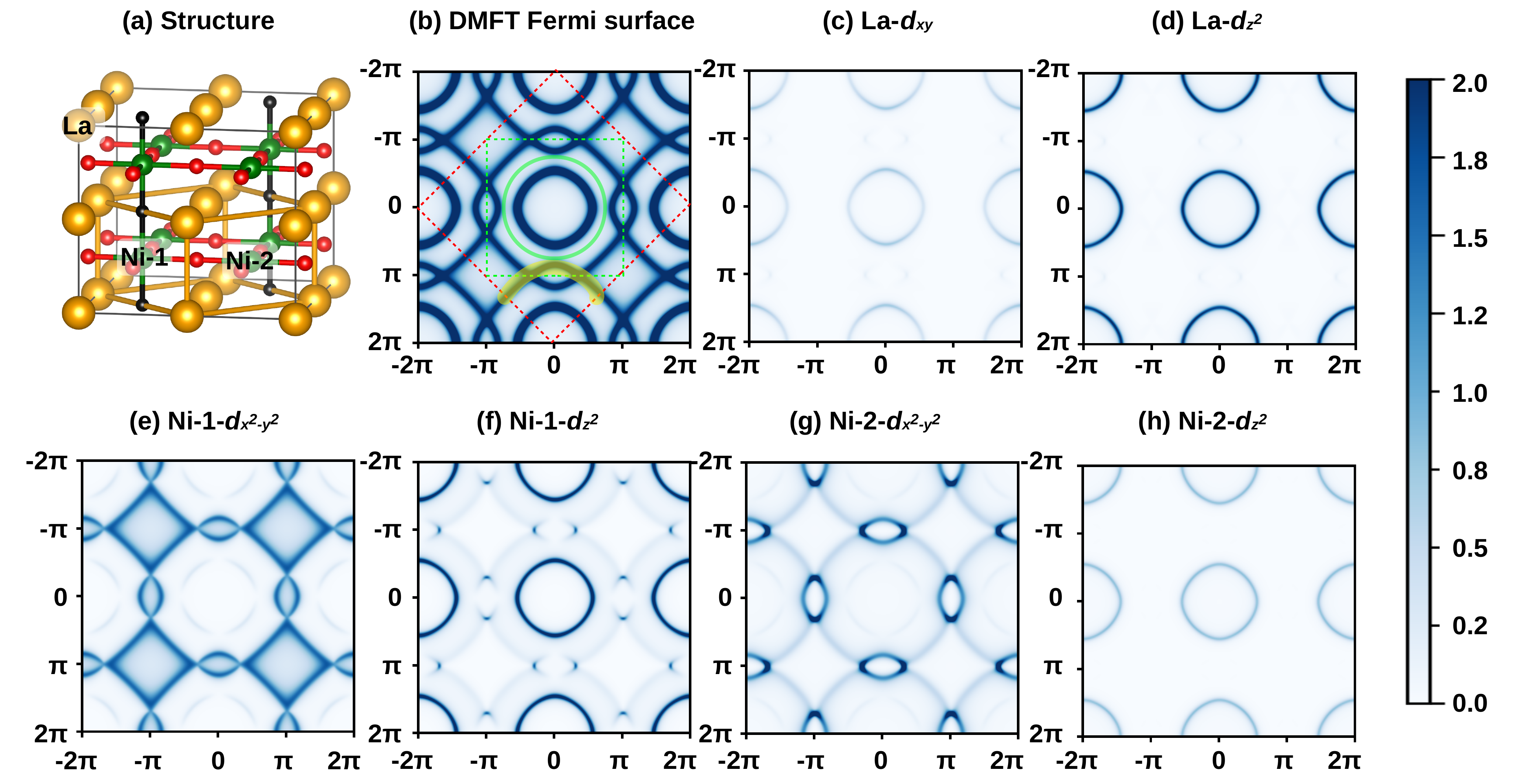}
\caption{DMFT Fermi surface (FS) of LaNiO$_2$H$_{\delta}$ ($\delta$=50\%) at $k_z=0$ (for other $k_z$ momenta, see Appendix~III Fig.~\ref{Fig9_dmft}). (a) Crystal structure and definition of Ni-1 and Ni-2 sites, the $\sqrt2\times\sqrt2\times$1 supercell indicated by the La-La bonds (yellow bars) is the computational model with inequivalent Ni-1 and Ni-2 site.
(b) Total Fermi surface. (c-h) Site- and orbital-resolved Fermi surface of La, Ni-1 and Ni-2. All Fermi surfaces are plotted in an extended Brillouin zone (BZ) for clarity. In (b) the dashed green square is the first BZ of the computational model as the crystal sublattice contains only two equivalent Nd and two non-equivalent Ni sites, while the dashed red square is the BZ that enables a comparison with the FS of the LaNiO$_2$ unit cell. The solid green circle encircles the $\Gamma$-pocket and the yellow shadow labels one quarter piece of \ls{the} Ni-$d_{x^2-y^2}$ FS.
}
\label{Fig5_H4_FS_DMFT}
\end{figure*}

In this subsection, we compute the Fermi surface of LaNiO$_2$H$_{\delta}$ and choose $\delta$=50\% as it leaves only one inequivalent Nd and two inequivalent Ni sites for the computationally heavy DMFT calculations. As shown in Fig.~\ref{Fig5_H4_FS_DMFT}(a), the normal Ni is again defined as ``Ni-2'' while the one between topotactic-H as ``Ni-1''.

As one can see in Fig.~\ref{Fig5_H4_FS_DMFT}(b), the FS of LaNiO$_2$H$_{0.5}$ not only contains the single-band $d_{x^2-y^2}$ character [labelled by the yellow area in Fig.~\ref{Fig5_H4_FS_DMFT}(b)], but also hosts an enhanced $\Gamma$-pocket. Please note that the original $A$-pocket of undoped LaNiO$_2$ now seems absent. This is because employing a $\sqrt{2}\times\sqrt{2}\times$1 supercell in computations rotates the BZ by 45$^{\circ}$, and shrinks its volume [see Fig.~\ref{Fig5_H4_FS_DMFT}(a,b) for details], which leads to the overlap of the $A$- and $\Gamma$-pocket.
Fig.~\ref{Fig5_H4_FS_DMFT}(d,f) indicates that the $\Gamma$-pocket is mainly constructed from La-$d_{z^2}$ and Ni-1-$d_{z^2}$. The former contribution is the same as in pure La(Nd)NiO$_2$ while the latter one from Ni-1-$d_{z^2}$ \ls{is affected by} the formation of H-chains. Besides La-$d_{z^2}$ and Ni-1-$d_{z^2}$, Ni-2-$d_{z^2}$ also plays a tiny role at the $\Gamma$-pocket, which originates from the hybridization between La(Nd)-$d_{z^2}$ and Ni-$d_{z^2}$. Another notable feature is the hybridization between Ni-1-$d_{x^2-y^2}$ and Ni-1-$d_{z^2}$. The complex reconstruction of the Fermi surface caused by the H-chains will also affect the Hall coefficient. For the $k_z$-dependence of the DMFT FS see Appendix~III and Fig.~\ref{Fig9_dmft}.

\subsection{F.~Magnetic properties and electronic correlations of LaNiO$_2$H$_{\delta}$}

In order to study the modifications of magnetic properties due to topotactic-H, we define and compute local (impurity) spin-spin correlation functions $\chi(\tau)=\sum_{mn}\langle S_z^m(\tau)S_z^n(0)\rangle$ of LaNiO$_2$ with different concentrations of topotactic-H using DFT+DMFT. Here, $S_z^m(\tau)$ is the $z$ component of the spin operator of orbital $m$ and $\tau$ the imaginary time.
\ls{The susceptibility} $\chi$ can generally be divided into orbital-diagonal (when~$m$=$n$) and off-diagonal contributions (when $m$$\neq$$n$) for all interacting sites.

\begin{figure*}[t]
\includegraphics[width=1.0\textwidth]{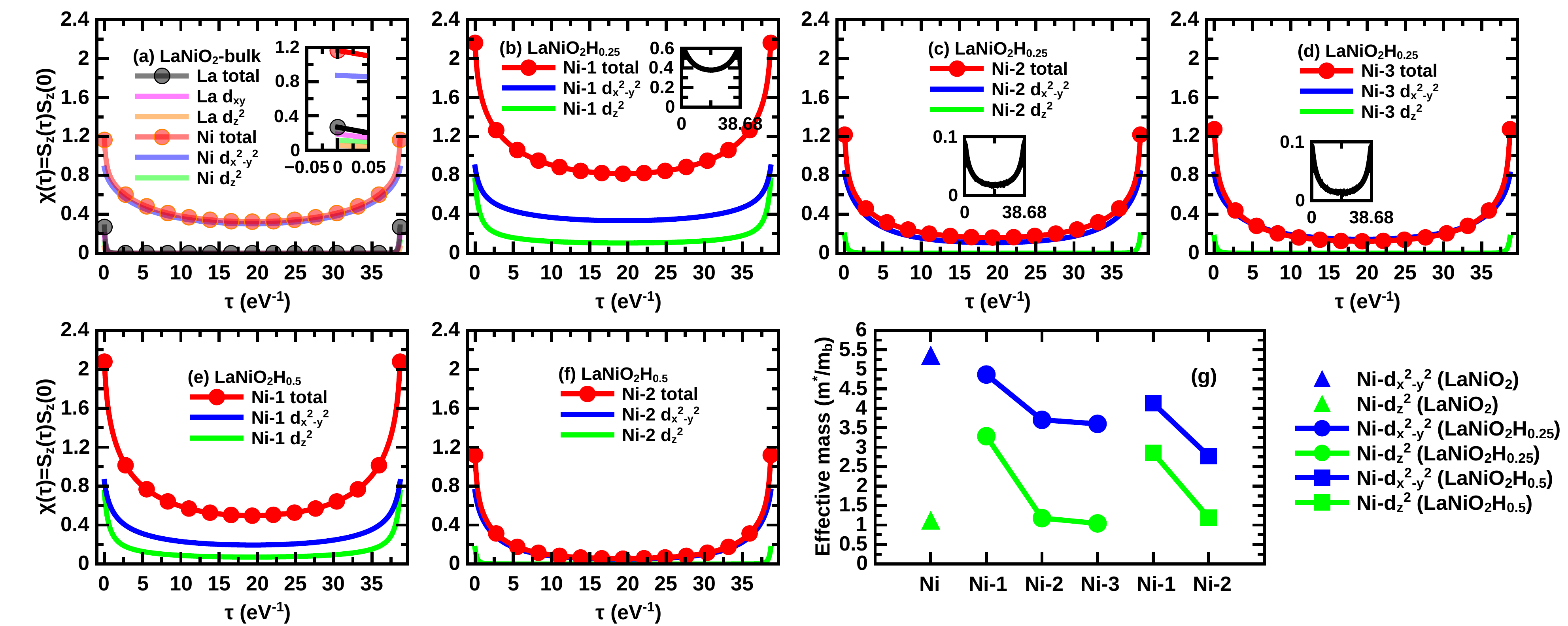}
\caption{DMFT site- and orbital-resolved local spin-spin correlation functions $\chi$($\tau$)=$<$S$_z$($\tau$)S$_z$(0)$>$ for LaNiO$_2$H$_{\delta}$ at \ls{$\beta$=38.68\,eV$^{-1}$}.
(a) bulk LaNiO$_2$ ($\delta$=0), the inset shows the zoom-in at $\tau$ \ls{from} -0.05 to 0.05\,eV$^{-1}$ for a clear observation;  (b-d)  LaNiO$_2$H$_{0.25}$ ($\delta$=25\%) for the three inequivalent Ni sites. Insets: additional interorbital (off-diagonal) contributions; (e,f)   LaNiO$_2$H$_{0.5}$ ($\delta$=50\%) for the two inequivalent Ni sites. In (a-f), the Ni-1, Ni-2 and Ni-3 indicate the first, second, and third nearest neighbor Ni site from the H-chains; (g) effective mass $m^{*}$/$m_{b}$ of the $d_{x^2-y^2}$ and $d_{z^2}$ orbitals of Ni-1 to Ni-3 sites in LaNiO$_2$,
LaNiO$_2$H$_{0.25}$ and LaNiO$_2$H$_{0.5}$, $m^{*}$ and $m_{b}$ indicate  effective mass from DMFT and DFT band mass, respectively.}
\label{Fig6_DMFT}
\end{figure*}

Furthermore, the value of $\chi$($\tau$) at $\tau$=0 can be interpreted in terms of the instantaneous local magnetic moment. As we again employ a La-5$d$+Ni-3$d$ Kanamori-type multi-site and multi-orbital model, it is instructive to study the importance of topotactic-H and La(Nd)-site. 
Fig.~\ref{Fig6_DMFT}(a) shows the results for bulk LaNiO$_2$ without topotactic-H. As one can see, the instantaneous moment of La at $\tau$=0 is merely 0.268\,$\mu_B^2$, and as indicated by the inset zoom-in of the orbital-resolved contributions, it originates mostly from the La-$d_{xy}$ and La-$d_{z^2}$ orbitals, which correspond to the contribution from the $A$-pocket and the $\Gamma$-pocket, respectively.  The moment of Ni \ls{(equal time susceptibility)} is 1.163\,$\mu_B^2$ at $\tau$=0\ls{,} and \ls{the susceptibility} decays to 0.323\,$\mu_B^2$ at $\tau$=$\beta$/2.
The magnetic moment of Ni can be attributed predominately to the Ni-$d_{x^2-y^2}$ orbital, indicating the essentially single-band nature of nickelate superconductors.
Other orbitals that are not shown, including Ni-$t_{2g}$ and La-$d_{yz/xz}$ and La-$d_{x^2-y^2}$ orbitals, have basically no or only negligible contributions. 

The observed fast decay of $\chi$ for both La and Ni at finite $\tau$ reflects a dynamical screening and strong damping of the local moment and may contribute to the absence of long-range magnetism in (undoped) infinite-layer nickelates. 
This fast screening is a consequence of self-doping and charge transfer away from the Ni-$d_{x^2-y^2}$ orbital even without Sr(hole)-doping \cite{chen2022theory}.

Fig.~\ref{Fig6_DMFT}(b-d) shows $\chi$($\tau$) of LaNiO$_2$H$_{0.25}$, which corresponds to a single H-chain in a 2$\times$2$\times$2 supercell as shown in Fig.~\ref{Fig1}(c). Such a configuration leads to three inequivalent Ni sites; Ni-1, Ni-2, and Ni-3 denote the first, second and third  nearest Ni neighbor from the H-defect. As shown in Fig.~\ref{Fig6_DMFT}(b), the moment of Ni-1 is remarkably enhanced from 1.163\,$\mu_B^2$ in LaNiO$_2$ to 2.161\,$\mu_B^2$ in Ni-1 of LaNiO$_2$H$_{0.25}$. \ls{The local susceptibility} $\chi$($\tau$) of Ni-2 and Ni-3 are, on the other side, however, \ls{is} basically identical to the Ni site in LaNiO$_2$ [Fig.~\ref{Fig6_DMFT}(a)]. Orbital-diagonal contributions from the $d_{x^2-y^2}$ and $d_{z^2}$ orbitals of Ni-1 do not fully account for the magnetic moment enhancement. In fact, they only show a finite enhancement of the $d_{z^2}$ contribution while the $d_{x^2-y^2}$ contribution remains unchanged. To identify the origin of such an enhanced moment, we further compute the site-resolved off-diagonal contribution for Ni-1 to Ni-3, as shown in the insets of Fig.~\ref{Fig6_DMFT}(b-d). This off-diagonal contribution is similarly large as the orbital-diagonal contribution of the $d_{z^2}$ orbital and explains the missing contribution to the magnetic moment.
This enhanced off-diagonal contributions of the instantaneous moment in Ni-1 is a consequence of Hund's exchange $J$, the reduced occupation, and enhanced correlations of the Ni-1 $d_{z^2}$-orbital. The boosted local magnetic moment is relatively robust against dynamical screening compared with the systems without H \ls{intercalation}. Since H-chains only affect the first nearest neighbor Ni sites between H atoms, such an effect is not observed in Ni-2 and Ni-3.

The results of $\chi$($\tau$) for LaNiO$_2$H$_{0.5}$ which corresponds to a structure with two H-chains arranged along the (110)-direction in a 2$\times$2$\times$2 LaNiO$_2$ supercell [Fig.~\ref{Fig1}(e)] are shown in Fig.~\ref{Fig6_DMFT}(e,f). As one can see Ni-1 is basically identical with Ni-1 in LaNiO$_2$H$_{0.25}$ [Fig.~\ref{Fig6_DMFT}(b)] and Ni-2 is basically identical with Ni in LaNiO$_2$ [Fig.~\ref{Fig6_DMFT}(a)] and Ni-2/Ni-3 in LaNiO$_2$H$_{0.25}$ [Fig.~\ref{Fig6_DMFT}(c,d)]. This shows us that not only with respect to the occupation but also with respect to the local magnetic moment, only the Ni sites next to the H-chain are modified.

Finally, we study the degree of electronic correlations in undoped LaNiO$_2$, LaNiO$_2$H$_{0.25}$ and LaNiO$_2$H$_{0.5}$, as shown in Fig.~\ref{Fig6_DMFT}(g). The effective mass of the $d_{x^2-y^2}$ and $d_{z^2}$ orbital in LaNiO$_2$ is 5.35 and 1.12, respectively. That is, the $d_{x^2-y^2}$ orbital is strongly correlated as it is nearly half-filled, while the $d_{z^2}$ orbital is essentially uncorrelated as it is basically fully filled. For LaNiO$_2$H$_{0.25}$ [that hosts one H-chain as shown in Fig.~\ref{Fig1}(c)], the correlations of the $d_{z^2}$ orbital at Ni-1 site are  boosted as indicated by a mass enhancement $m^*$/$m_b$=3.28. On the contrary, and consistent with the previous discussion, $m^*$/$m_b$ of the $d_{z^2}$ orbitals of Ni-2 and Ni-3 remain close to one, i.e., 1.17 and 1.04, respectively, as this orbital is almost completely filled  and essentially the same as for the defect-free compound.
This again indicates that topotactic-H only affects the nearest-neighbor Ni site. The mass enhancement of the $d_{x^2-y^2}$ orbitals of Ni-1 to Ni-3 in LaNiO$_2$H$_{0.25}$ shows a similar trend as that of the $d_{z^2}$ orbital: Ni-1 $d_{x^2-y^2}$ is the most correlated one, while correlations at the Ni-3 site are the weakest. Correlations of all the $d_{x^2-y^2}$ orbitals are weaker for compounds with H-chains compared to LaNiO$_2$. This is because we are further away from half-filling: 
\ls{As shown in Table~\ref{Tab:occupation} for LaNiO$_2$H$_{0.5}$, topotactic-H absorbs electrons mostly from the $d_{z^2}$ orbital of the nearest Ni-1 ion, with the second largest (though already much smaller) contribution stemming from the $d_{x^2-y^2}$ orbital of Ni-2. Consequently, the occupation $n$ of Ni-1 $d_{z^2}$ is reduced from full-filling ($\sim$2/orbital) to 1.11/orbital and $n$ of Ni-2 $d_{x^2-y^2}$ orbital is reduced from half-filling ($\sim$1/orbital) to 0.83/orbital. As a result, the correlations at Ni-1 $d_{z^2}$ and Ni-2 $d_{x^2-y^2}$ are increased and reduced, respectively.}
Let us reiterate here that low-density H-defects in LaNiO$_2$H$_{\delta}$ alter valence, magnetic moment, and the degree of correlations mainly on the closest Ni site while leaving other sites basically unchanged. This might explain the coexistence of different valence states in experiment \cite{krieger2021charge} when such defects are present. Moreover, previous work focusing on the critical temperature $T_C$ of NdNiO$_2$ demonstrated that correlations are decisive for $T_C$ \cite{Kitatani2020}. The additional $d_{z^2}$ Fermi surface for the doped compound is expected to be unfavorable for superconductivity.
Hence our results may explain why $T_C$ in nickelates varies more strongly from sample to sample and the transition is generally broader than in other superconductors \cite{li2019superconductivity,Li2020,zeng2020}. 

\section{IV.~Conclusions and Outlook}
Our DFT and DMFT  computations demonstrate the general tendency to form 1D H-chains in the infinite-layer nickelate LaNiO$_2$ if excess hydrogen is available, e.g., from the CaH$_2$ reducing agent \ls{or atmospheric humidity}. The topotactic-H atoms are confined in the $z$-direction (001) if LaNiO$_2$ is grown on a SrTiO$_3$ substrate, \ls{as} the formation of  Ni-$d_{z^2}$-H-1$s$ bonds \ls{is energetically profitable}. If the excess hydrogen exceeds a single chain, additional chains are formed at neighboring sites in the $xy$-direction (110) due to the La-$d_{xy}$-H-1$s$ $\delta$-type bonds. 
The existence of H near the boundary of infinite-layers nickelates was confirmed by Ref.~\cite{Puphal2022,Choi2020}, and its existence in deeper parts \ls{of the sample} is worth to be investigated further. In Ref.~\cite{si2022fingerprints} we proposed flat phonon vibrations near $\sim$103 and $\sim$178\,meV as possible indications for topotactic H, which is a key characteristic as both of these peaks are located at energies that are perfectly separated from the \ls{other transition peaks such as orbital excitations within the Ni-3$d$ orbitals ($\sim$1.3\,eV) and other phonons ($\leq$0.07\,eV) \cite{Lu2021}. However, their proximity to the magnon excitation ($\sim$0.2\,eV) or a possible spectral feature due to Ni-3$d$ and Nd-5$d$ orbital hybridization ($\sim$0.75\,eV) makes them harder, but not impossible, to detect in RIXS.}
Such 1D H-chains can be expected to result in several uncertainties regarding the following aspects in infinite-layer nickelate superconductors: (1) structural distortion; (2) magnetism; (3) coexistence of various oxidation states of Ni ions\ls{, and (4) charge ordering. These four points are discussed in more detail in the following:}

More specifically,
(1) our phonon computations reveal that LaNiO$_2$H$_{\delta}$ is dynamically stable, and  DFT structural relaxations predict a lattice expansion as $\delta$ increases from 0\% (LaNiO$_2$) to 100\% (LaNiO$_2$H) (see Appendix~IV and Fig.~\ref{FigB5}). In realistic samples, the topotactic-H may distribute inhomogeneously. Hence, this lattice expansion effect may be rather local, leading to non-zero internal stress near individual H-chains. 
The experimentally observed ``shear fault'' phase \cite{zeng2020,lee2022character} is expected as an effective way to eliminate internal stress, i.e., when the internal stress is accumulated and exceeds a threshold value, ``shear fault'' is generated by removing a Nd(La)O layer once along the $z$-direction and a NiO$_2$ layer once along the $xy$-plane to release internal stress. In fact, one notable feature of the experimentally observed ``shear fault'' \cite{zeng2020,lee2022character} is its rectangle shape along the $z$-direction, which hints at a 1D character of the driving force or defect details. H-chains, which are difficult to detect,  might explain this.

(2) The general experimental consensus is an absence of long-range magnetic order in pure and doped La(Nd)NiO$_2$ \cite{PhysRevResearch.4.023093}, which differs from cuprates and several theoretical computations that yield $C$- \cite{bernardini2020infinite,PhysRevB.102.205130} or $G$-type \cite{Siheon2019} AFM order as energetically stable compared with FM or $A$-AFM order.
Observations supporting short-range AFM correlations have been reported by nuclear magnetic resonance spectroscopy (NMR) measurement that indicates local AFM spin fluctuations in (Nd,Sr)NiO$_2$ \cite{Cui2021} while the presence of FM-like short-range order was supported by the slow magnetically dynamical process under external strong field \cite{PhysRevResearch.4.023093}. Additionally, spin-glassy and spin-freezing behavior, which are commonly induced by local geometric/magnetic frustrations, have been observed in both infinite- \cite{lin2022universal} and finite-layer nickelates \cite{PhysRevB.102.054423} through powder neutron diffraction, magnetization, and muon-spin rotation ($\mu$SR) measurements. Ref.~\cite{zhou2022antiferromagnetism} reported the coexistence of FM and AFM regions and short-range AFM order by observing large exchange bias effects and performing magnetic linear dichroism measurements, respectively. Another strong evidence for a sizable AFM exchange interaction between Ni sites is the direct observation of a dispersive magnetic excitation that shares some common features with spin waves in spin-1/2 AFM square lattice in cuprates \cite{PhysRevLett.105.157006,PhysRevLett.117.237203,PhysRevLett.105.247001}.
While the absence of AFM order may be explained by the self-doping of the Ni $d_{x^2-y^2}$ orbital even for the parent compound due to the La(Nd) pockets, topotactic-H can further diversify the picture and explain the quite different experimental observations.

Our spin-polarized DFT and DMFT spin-spin correlation $\chi(\tau)$ computations demonstrate that the formation of H-chains triggers a 2D intra-layer AFM to 3D AFM transition, with enhanced inter-layer coupling. Both the spin-spin correlations of the total, $d_{x^2-y^2}$ and $d_{z^2}$ components are boosted by the nearby topotactic-H. The enhanced instantaneous magnetic moment and competing magnetic orders can be expected to contribute to the formation of local magnetic frustration, leading to the observed spin-glass and/or spin-freezing behavior \cite{lin2022universal}.

(3) Ni$^{2+}$ (3$d^8$) oxidation states have been reported in a recent study \cite{krieger2021charge} and our previous research \cite{Si2019} was recalled to explain the origin of Ni$^{2+}$. However, in their XAS spectra, a Nd-Ni hybridization was observed, which is in contradiction to the theoretical results of LaNiO$_2$H, where the La(Nd)-Ni hybridization is fully eliminated by H-defects \cite{Si2019}. In the present paper, we resolve this apparent contradiction: the formation of H-chains naturally leads to coexisting Ni$^{1+}$ and Ni$^{2+}$ states. Indeed, any topotactic-H concentration below 100\% (pure LaNiO$_2$H) is able to split the Ni atoms into two subgroups: nominal $S$=1/2 Ni$^{1+}$ (3$d^9$) and unusual $S$=1 Ni$^{2+}$ (3$d^8$), \ls{while} keeping the La-Ni hybridization. The degree of La-Ni hybridization is tuned by and anti-proportional to the topotactic-H density. 

\ls{(4) We find the largest topotactic-H $E_b$ if H-chains form along the $z$-axis and are arranged in such a way that H-chains neighbor in the direction along the $xy$-diagonal but have a distance of three unit cells at low H-concentration $\delta$$\sim$10\% (two unit cells at high H-concentration $\delta$$\sim$50\%) to the next H-chain in both the $x$- and $y$-direction.
As this also encompasses a corresponding  alternation in the Ni oxidation states \footnote{Through the Nd-Ni hybridization, this will further result in a charge ordering pattern of the Nd sites.}, it possibly explains the recently observed charge order state in nickelate superconductors \cite{krieger2021charge,tam2021charge}.}

Finally, our predictions of 1D H-chains in infinite-layer nickelates may explain the previous experimental observations regarding uncertainties of their electronic structures, magnetic orders, and lastly, the hidden superconducting mechanism and pairing interactions. We expect the areas with stoichiometric LaNiO$_2$ to exhibit  purely $d$-wave superconductivity and AFM fluctuations. The H-polluted regions are expected to exhibit magnetic and structural frustration as well as a suppression of superconductivity. Currently, the difficulties of identifying such H-chains in experiments stem from: (1) the radius and mass of H are negligible compared with those of La(Nd), Ni, and O. This makes them hard to \ls{detect} by commonly employed techniques such as X-ray powder diffraction (XRD) or scanning transmission electron microscopy (STEM). (2) As revealed by the above phonon computations, the dynamical stability of La(Nd)NiO$_2$ does not rely on the concentration of topotactic-H atoms. Hence, the $AB$O$_2$ infinite-layer structure should be detected by STEM even in the presence of H. (3) As revealed by the above presented DMFT spectra and spin-correlations functions (instantaneous magnetic moment), topotactic-H atoms merely affect the closest Ni atoms by absorbing one electron from the Ni-$d_{z^2}$ orbital and forming La-$d_{xy}$-H-1$s$ and Ni-$d_{z^2}$-H-1$s$ $\delta$-bonds. Hence further x-ray spectra, \ls{infrared optical} spectroscopy and precise chemical measurements are worth performing in order to achieve a final conclusion on the existence of diffused H-defects or chains.

\section{Acknowledgments}

\begin{acknowledgments}
We thank Z.~Zhong, M.~Kitatani, R.~Arita, O.~Janson, J.~M.~Tomczak and S.~D.~Cataldo for helpful comments and discussions.
We acknowledge funding through the Austrian Science Funds (FWF) project I 5398.
L.~S. thanks the starting funds from Northwest University.
Calculations have been done on the Vienna Scientific Cluster (VSC).
\end{acknowledgments}

\section{Appendix}

\subsection{I.~Ground state structures of LaNiO$_2$H$_{\delta}$}

\begin{figure*}
\includegraphics[width=0.8\textwidth]{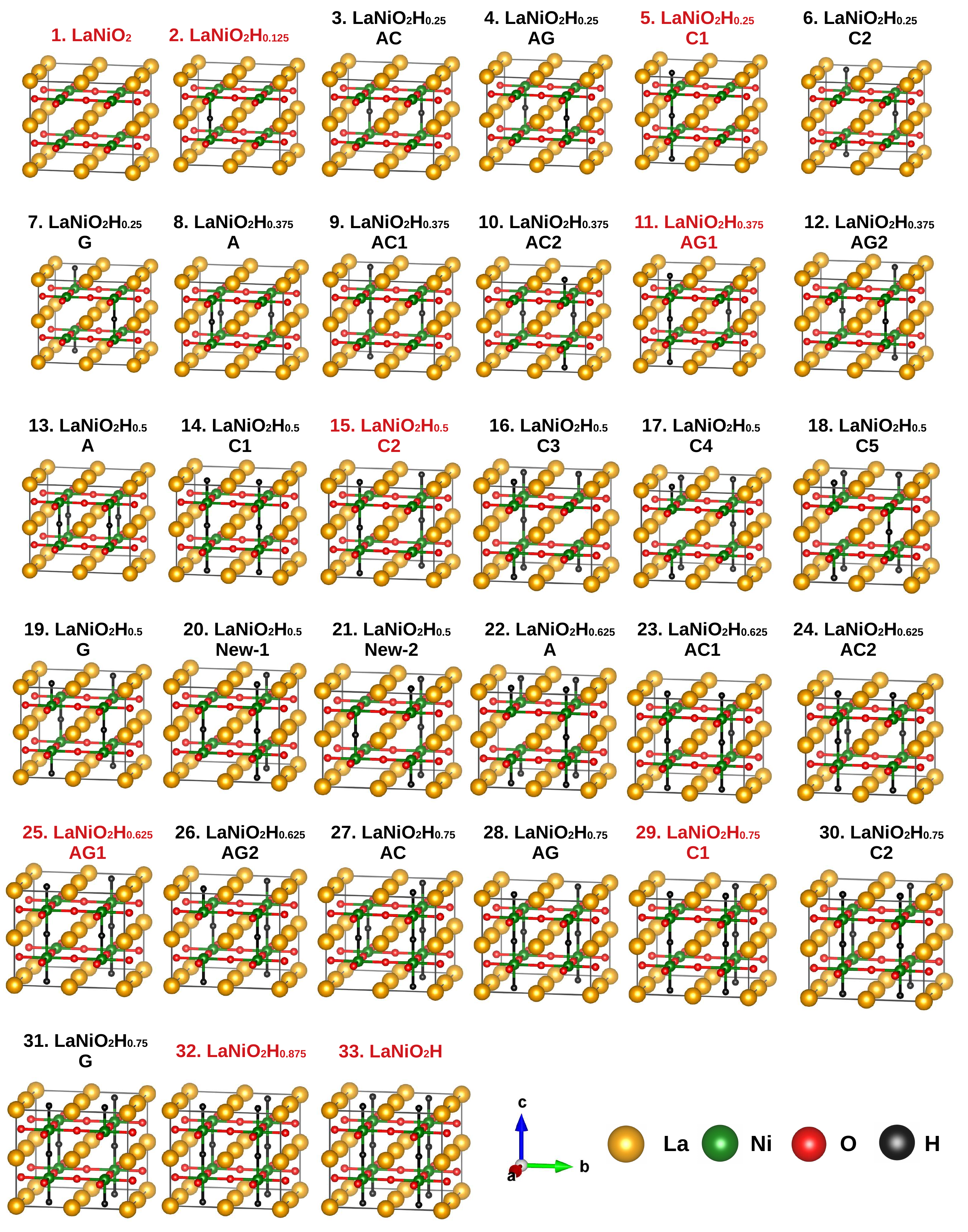}
\caption{All possible inequivalent combinations of apical hydrogen (H) distribution in a 2$\times$2$\times$2 supercell of LaNiO$_2$H$_{\delta}$ for $\delta$=0.0\%, 12.5\%, 25\% , 37.5\%, 50\%, 62.5\%, 75\%, 87.5\% and 100\%. The structures labeled by red texts are the corresponding ground states with the lowest DFT total energy for each concentration.}
\label{Fig7_Structure}
\end{figure*}

\begin{figure}
\includegraphics[width=0.38\textwidth]{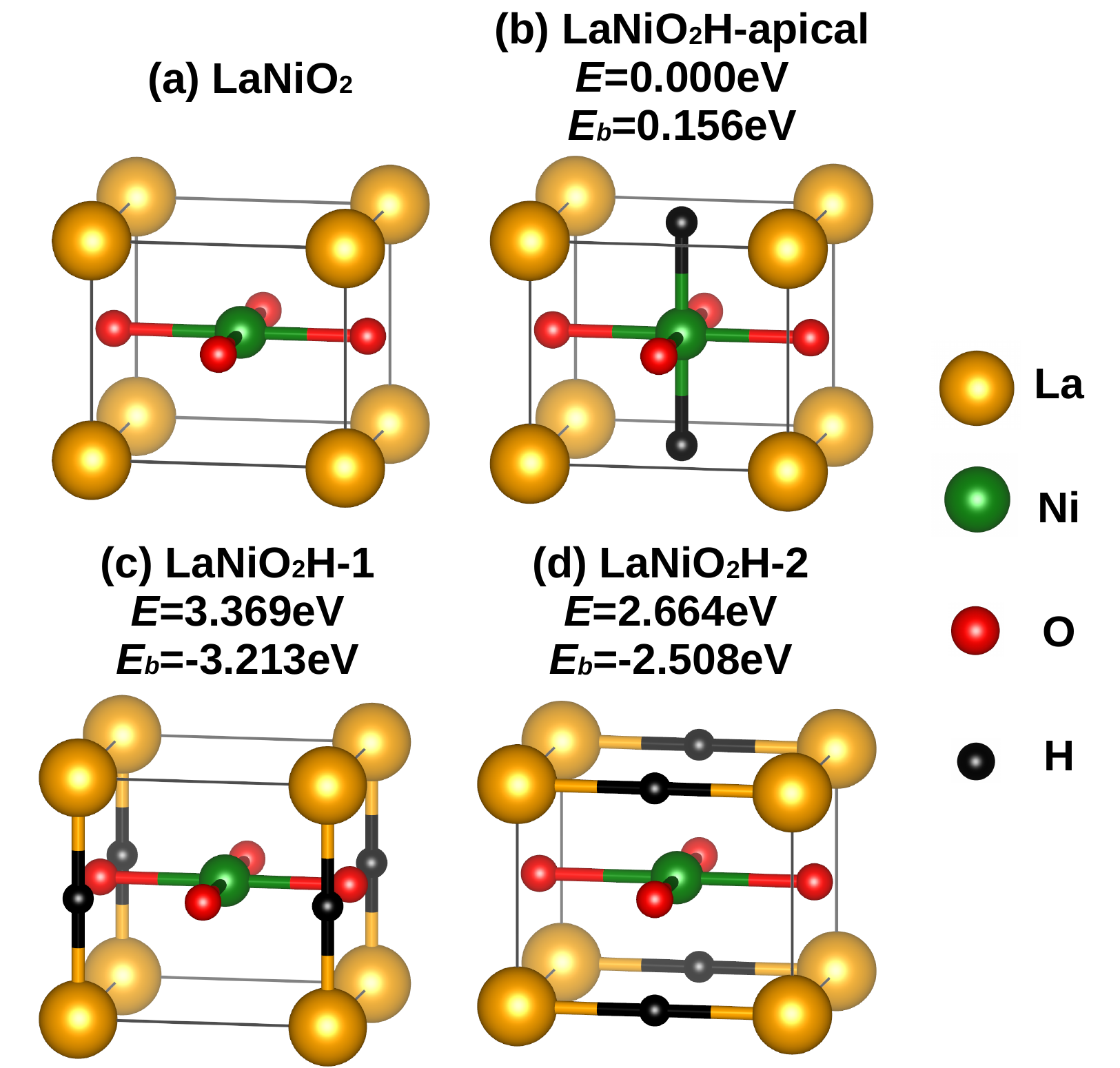}
\caption{(a) Stoichiometric LaNiO$_2$; (b-d) LaNiO$_2$H with topotactic-H occupying the out-of-plane apical position (b), in-plane position,  and (c) the out-of-plane position between La-La bond (d). The $E$ are the total energies of the different topotactic-H configurations in (b,c,d) and the total energy ($E$) of LaNiO$_2$H with apical topotactic-H (b) is set as zero for comparison. The $E_b$'s are binding energies of topotactic-H.}
\label{FigA2-1}
\end{figure}

\begin{figure}
\includegraphics[width=0.45\textwidth]{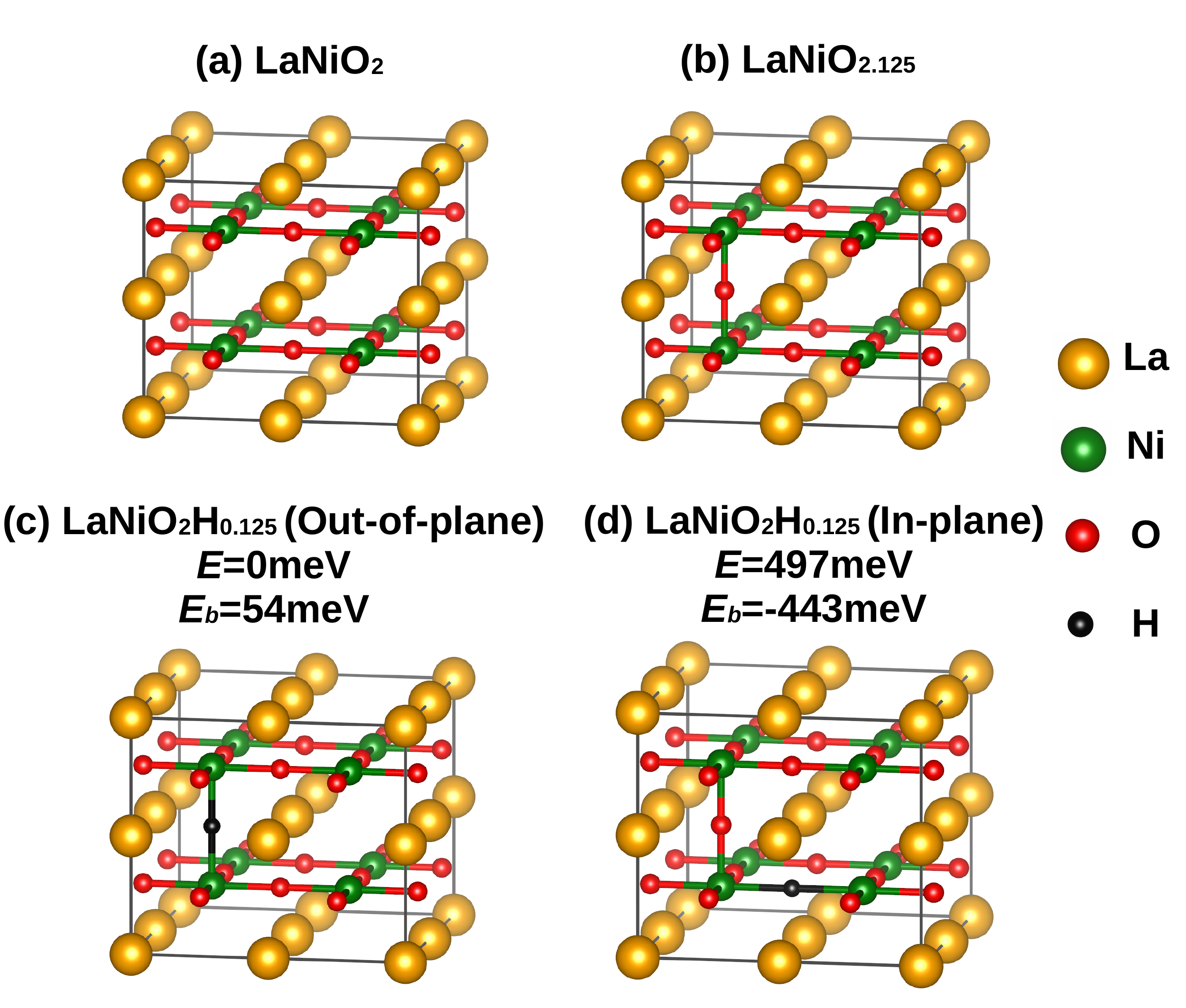}
\caption{(a) 2$\times$2$\times$2 stoichiometric LaNiO$_2$; (b) LaNiO$_2$H supercell with a single additional O atom; (c) single topotactic-H atom replace the out-of-plane (apical) position O and (d) the in-plane O. 
The $E$ are the total energies of the different topotactic-H configurations in (c,d), the total energy ($E$) of LaNiO$_2$H with apical topotactic-H (c) is set as zero for comparison. The $E_b$'s are binding energies of topotactic-H.}
\label{FigA2-2}
\end{figure}

\ls{Additional DFT calculations for the crystal structures of LaNiO$_2$H$_{\delta}$ are shown in Fig.~\ref{Fig7_Structure}, \ref{FigA2-1} and \ref{FigA2-2}.
We have analyzed all of the possible apical H positions of LaNiO$_2$H$_{\delta}$ (0.0\%$<$$\delta$$<$100\%) shown in Fig.~\ref{Fig7_Structure}. Their total energies as obtained by DFT calculations as shown in Table~\ref{Tab:energy}. 
This supplements Fig.~\ref{Fig1} of the main text, where only the stable structures of the selected concentrations of topotactic-H are presented.
Please note that for LaNiO$_2$, LaNiO$_2$H$_{0.125}$, LaNiO$_2$H$_{0.875}$ and LaNiO$_2$H there is only one in-equivalent H position. In all the considered structures, the NiO$_2$ planers are preserved without removing the O atoms. This assumption is based on our DFT calculations as shown in Fig.~\ref{FigA2-1} and Fig.~\ref{FigA2-2}. 

Let us now discuss which position in the unit cell is most favorable for the topotactic-H to occupy by considering two scenarios:
(1) First, we consider stoichiometric LaNiO$_2$ and would like to know which location will an intercalated hydrogen occupy. This question has already been addressed in our previous study \cite{Si2019} and the answer is that the topotactic-H occupies the out-of-plane (apical) position above/under the Ni ions and not between the La ions, neither in-plane nor out-ot-plane (which would lead to a La-H-La bond).
Fig.~\ref{FigA2-1}(a) shows stoichiometric LaNiO$_2$, Fig.~\ref{FigA2-1}(b) shows the case of topotactic-H at the apical position between Ni atoms, Fig.~\ref{FigA2-1}(c) shows in-plane topotactic-H and Fig.~\ref{FigA2-1}(d) shows the topotactic-H between La ions. Our DFT computations indicate that  LaNiO$_2$H with apical position H has the lowest energy compared to the in-plane and La-H-La case in Fig.~\ref{FigA2-1}(c,d). The energies of the structures in Fig.~\ref{FigA2-1}(c,d) are 3.369\,eV and 2.664\,eV per LaNiO$_2$H higher than the apical topotactic-H case in Fig.~\ref{FigA2-1}(b). We additionally compute the topotactic-H binding energy ($E_b$) of the three cases, for apical topotactic-H $E_b$ is 0.156\,eV while $E_b$ of the in-plane and the La-H-La case is -3.213\,eV and -2.508\,eV, respectively. These negative $E_b$ of Fig.~\ref{FigA2-1}(c,d) indicate that intercalating topotactic-H is actually energetically unfavorable for these two cases.

(2) Let us now consider a 2$\times$2$\times$2 supercell of LaNiO$_2$ plus a single O atom, which allows for the coexistence of in-plane and out-of-plane O atoms. Now we would like to know which O atom will be replaced by a topotactic-H. We show this situation in Fig.~\ref{FigA2-2}(a-d). Fig.~\ref{FigA2-2}(a) shows the undoped stoichiometric LaNiO$_2$ supercell, and in Fig.~\ref{FigA2-2}(b) there is an additional apical O occupying the empty position between the La layers, Fig.~\ref{FigA2-2}(c,d) reflects the topotactic process for the apical (out-of-plane) and in-plane O, respectively. As shown in Fig.~\ref{FigA2-2}(c,d), the out-of-plane topotactic-H at the apical position (c) is energetically favorable compared to the in-plane one (d) by 497\,meV per topotactic-H. Moreover, $E_b$ in Fig.~\ref{FigA2-2}(c) is positive (54\,meV/topotactic-H) while negative (-443\,meV/topotactic-H) in Fig.~\ref{FigA2-2}(d), indicating that H-intercalation is energetically favorable in Fig.~\ref{FigA2-2}(c) while unfavorable in Fig.~\ref{FigA2-2}(d).
 
The DFT calculations presented above reveal that in both situations (1-2), occupying the apical position above the Ni atoms is the energetically preferred location for intercalated H atoms, at least as for all topotactic positions of the original perovskite lattice structure.. As a result, we focus on topotactic-H at the (former) apical oxygen position in this paper.This will also be the location where topotactic hydrogen can be expected to reside in experimentally grown films.}

\begin{figure*}
\includegraphics[width=0.8\textwidth]{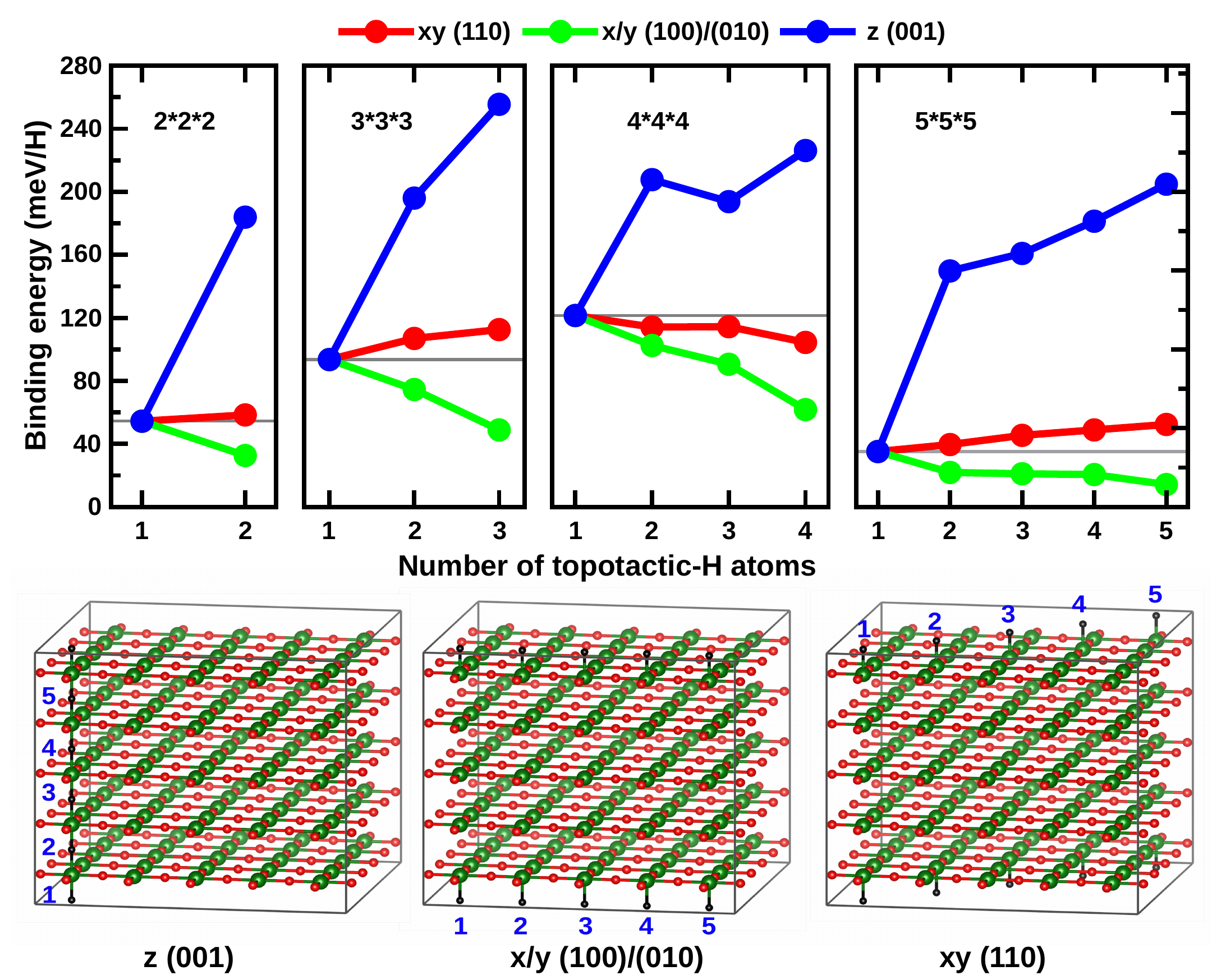}
\caption{$E_b$ per topotactic-H (top) in the formation process of 1-D H-chains along different directions of $z$ (001), $x$/$y$ (100)/(010) or $xy$ (110) (bottom). The constant grey line indicates the $E_b$ of single topotactic-H in each size supercell. The $x$-axis shows the number of topotactic-H atoms, indicating the formation process from single topotactic-H to finite-size H-chains and then to infinite H-chains.
Bottom: crystal structures of LaNiO$_2$ 5$\times$5$\times$5 supercell with additionally intercalated topotactic-H atoms along $z$ (001), $x$ or $y$ (100)/(010) and $xy$ (110) direction, respectively. For simplification, all La atoms are not shown. The Ni, O and H atoms are labeled as green, red and black balls. The blue numbers in the three structures represent the intercalating sites  where topotactic-H is gradually intercalated into the corresponding supercells  when calculating $E_b$ in the top panel.}
\label{FigA1}
\end{figure*}

\ls{In the current research, the formation process of 1D H-chains is therefore simulated by intercalating topotactic-H atoms into the apical positions in a 2$\times$2$\times$2 LaNiO$_2$ supercell. This guarantees a balance between computational efforts, accuracy and physical picture. Moreover, the recent experiment \cite{puphal2022synthesis} reports that the concentration of topotactic-H approaches $\sim$20\% in LaNiO$_2$ (i.e.~LaNiO$_2$H$_{0.2}$) and even approaches $\sim$100\% in the decomposed LaNiO$_{3-x}$H$_y$ sample (i.e.~LaNiO$_{3-x}$H). The former concentration $\sim$20\% (LaNiO$_2$H$_{0.2}$) basically reflects the situation when a single H-chain is formed in a 2$\times$2$\times$2 supercell of LaNiO$_2$, i.e., LaNiO$_2$H$_{0.25}$. This demonstrates the validity of our model. To prove that the formation of 1D H-chains is independent of the size of the employed supercell, we re-perform the key (non-spin-polarized DFT) calculations by employing 3$\times$3$\times$3, 4$\times$4$\times$4 and 5$\times$5$\times$5 LaNiO$_2$ supercells and intercalating topotactic-H atoms into the supercell along $x$-direction (100) [equal to $y$ (010)], $xy$-direction (110) and $z$-direction (001), respectively. 

The corresponding results are shown in Fig.~\ref{FigA1}. As a comparison, we also show the result of a 2$\times$2$\times$2 supercell, which was already discussed in the main text. Fig.~\ref{FigA1} top panel shows the corresponding $E_b$ per topotactic-H when $n \leq m$ ($n$ is the number of topotactic-H atoms and $m$$\times$$m$$\times$$m$ is the size of supercell), and the bottom panel shows the crystal structures when intercalating 1-5 topotactic-H atoms into the 5$\times$5$\times$5 LaNiO$_2$ supercell along $x$, $xy$, $z$ direction, respectively.

First, for all different supercells and all different H-chain lengths [both finite (when $n$$<$$m$) and infinite (when $n$=$m$)], $E_b$ of the (001) $z$-direction H-chain is higher than that of the H-chains in $x$/$y$- or $xy$-direction. Moreover, for all 1$<$$n$$\leq$m, the  $E_b$ of $z$-direction H-chain is always higher than that of single topotactic-H, indicating the formation of  (finite and infinite) 1D H-chains in the $z$-direction are not decided by the size of the supercell and concentration of topotactic-H. 

Second, the $E_b$ of the (001) $z$-direction infinite H-chains increases from 204.8\,meV/H in 5$\times$5$\times$5, to 255.5\,meV in 3$\times$3$\times$3, then decreases to  183.9\,meV in 2$\times$2$\times$2 and 158\,meV in fully topotactic LaNiO$_2$H \cite{Si2019}. The infinite 1D H-chain the 3$\times$3$\times$3 supercell hosts the largest $E_b$ of 255.5\,meV per topotactic-H.

Such behaviors can be understood by the competition between bonded topotactic-H and lattice expansion effect: when $\delta$ in LaNiO$_2$H$_{\delta}$ increases from zero to finite, on the one hand, the bonds formed between topotactic-H stabilize the topotactic region, leading to a larger $E_b$; on the other hand, the gradually increased topotactic-H  concentration results in a lattice expansion (see Appendix~IV and Fig.~\ref{FigB5}). This leads to local strain and enhances the total energy of the parent LaNiO$_2$ sublattice. 
As a consequence of these two balancing effects, there is a maximum $E_b$ per topotactic-H for 
$0<\delta<100\%$.

Please note that the energetically favorable infinite/finite H-chain in 3$\times$3$\times$3 supercell leads to structural and chemistry periodic with a vector of $R$=($\frac{1}{3}$, $\frac{1}{3}$, 0) along (110) $xy$-direction, this is consistent with the very recent experimentally observed charge/orbital order states vector in nickelate superconductors \cite{tam2021charge,krieger2021charge}.
}

\begin{table*}
\caption{DFT computed total energies in meV of LaNiO$_2$H$_{\delta}$ (LNO$_2$H$_{\delta}$: 12.5\%$<$$\delta$$<$87.5\%). For the concentrations ($\delta$=25\%, 37.5\%, 50\%, 62.5\%, 75\%) with more than one possible crystal structure (STR), the total energy of $A$ and $AC$ phases are set as zero. The structures and relative energies labeled with boldface are the ground states. The last two rows show $E_b$ (in units of meV) from non-spin-polarized DFT and spin-polarized DFT+$U$ calculations, respectively.}
\begin{tabular}{c|c|c|c|c|c|c|c|c}
\hline
\hline
$\delta$ (100\%)  & LNO$_2$H$_{0.125}$ & LNO$_2$H$_{0.25}$ & LNO$_2$H$_{0.375}$ & LNO$_2$H$_{0.5}$ & LNO$_2$H$_{0.625}$ & LNO$_2$H$_{0.75}$ & LNO$_2$H$_{0.825}$ & LNO$_2$H \\ 
\hline
STR-1 & 0.0 & AC (0.0)    & A (0.0)      & A (0.0)      & A (0.0) & AC (0.0) & 0.0 & 0.0 \\
\hline
STR-2 & - &  AG (-51.5)  & AC1 (-311.3) & C1 (-763.3)  & AC1 (-417.3) & AG (115.0) & -& - \\
\hline
STR-3 & - & $\mathbf {C1 (-302.5)}$ & AC2 (-15.0)   & $\mathbf {C2 (-924.9)} $ & AC2 (-16.7) & $\mathbf {C1 (-454.0)}$ & - & - \\
\hline
STR-4 & - & C2 (54.7)   & $\mathbf {AG1 (-449.1)}$  & C3 (-336.6)  & $\mathbf {AG1 (-438.2)}$ & C2 (60.3) & - & - \\
\hline
STR-5 & - & G (-96.9)   & AG2 (84.2)   & C4 (-436.3)  & AG2 (82.8) & G (83.5) & - & - \\
\hline
STR-6 & - &     -       &     -        & C5 (-1.4)    & - & - & - & - \\
\hline
STR-7 & - &     -       &     -        & G (114.3)    & - & - & - & - \\
\hline
STR-8 & - &     -       &     -        & New-1 (-385.3) & - & - & - & - \\
\hline
STR-9 & - &     -       &     -        & New-2 (-389.9) & - & - & - & - \\
\hline
\hline
$E_b$ (DFT) & 54.3 & 183.9 & 169.8 & 226.3 & 146.8 & 182.5 & 130.4 & 158.5 \\
\hline
$E_b$ (DFT+$U$) & 75.0 & 327.3 & 214.9 & 330.1 & 215.8 & 306.6 & 234.9 & 303.9 \\
\hline
\hline
\end{tabular}
\label{Tab:energy}
\end{table*}

\subsection{II.~Phonon of LaNiO$_2$H$_{\delta}$}

\begin{figure*}
\includegraphics[width=1.0\textwidth]{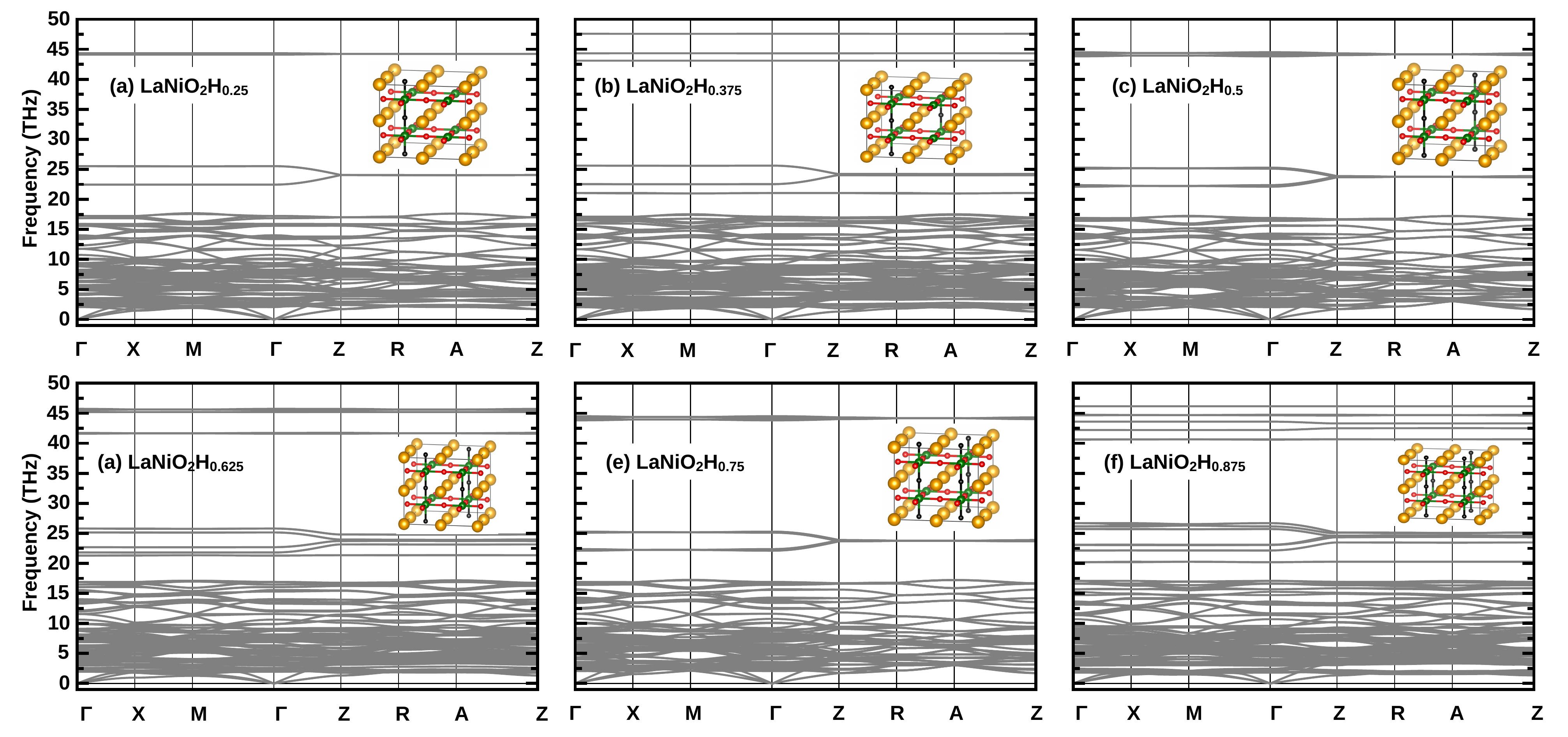}
\caption{DFT phonon spectra of LaNiO$_2$H$_{\delta}$ for $\delta$=25\% , 37.5\%, 50\%, 62.5\%, 75\% and 82.5\%. For results of $\delta$=0\% (LaNiO$_2$) and 100\% (LaNiO$_2$H) see main text Fig.~\ref{Fig3}, and for result for $\delta$=12.5\% (LaNiO$_2$H$_{0.125}$) see Ref.~\cite{si2022fingerprints}.}
\label{Fig8_phonon}
\end{figure*}

Fig.~\ref{Fig8_phonon} supplements Fig.~\ref{Fig3} and shows additional phonon spectra for various H concentrations. This demonstrates that the optical phonon modes induced by topotactic H around  $\sim$25 and 43\,THz are quite insensitive to the concentration of hydrogen or the chain formation.

\subsection{III.~$k_z$-dependence of Fermi surface of LaNiO$_2$H$_{\delta}$}

\begin{figure*}
\includegraphics[width=0.95\textwidth]{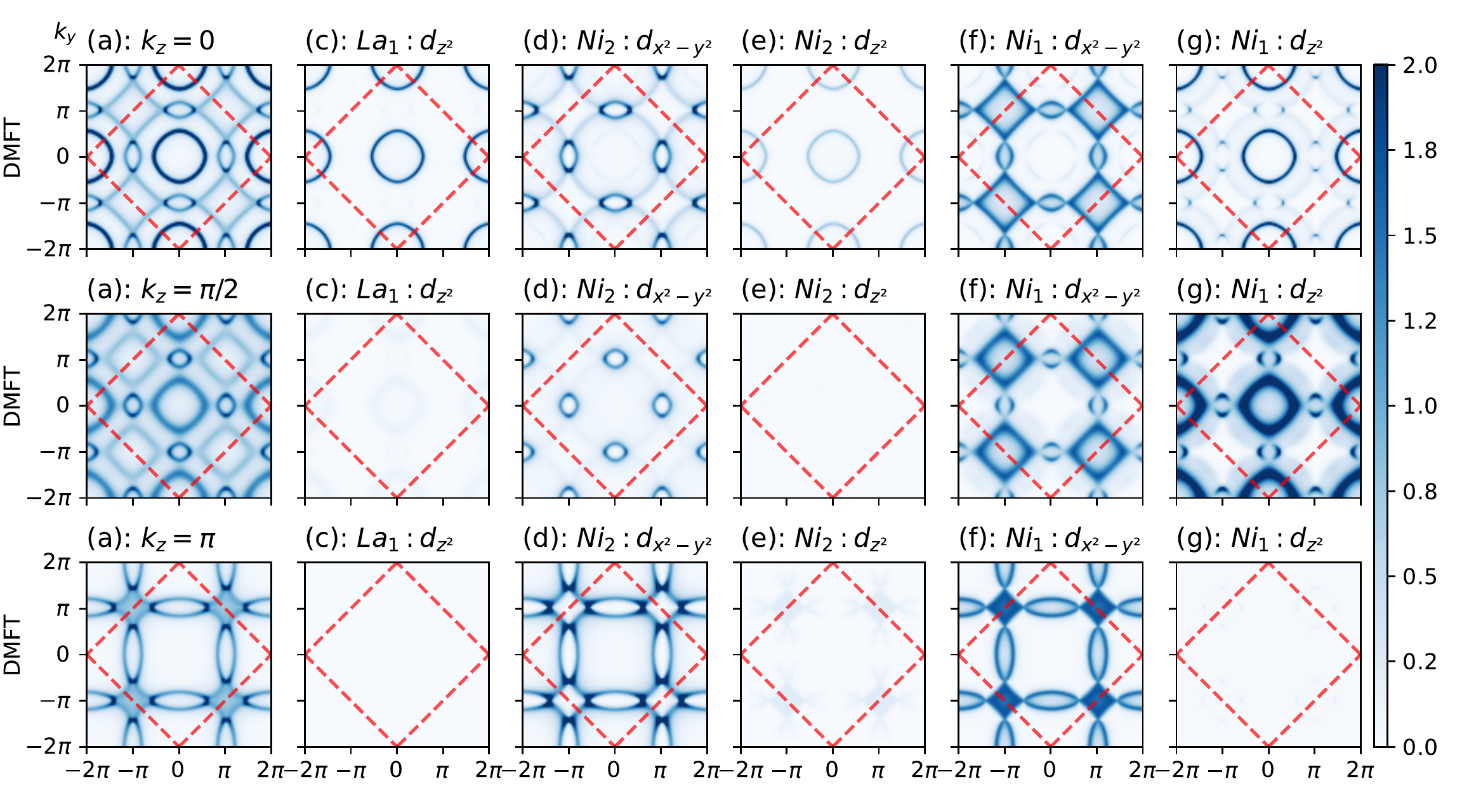}
\caption{DMFT Fermi surfaces for additional $k_z$'s besides the $k_z=0$ of Fig.~\ref{Fig5_H4_FS_DMFT} for LaNiO$_2$H$_{0.5}$.}
\label{Fig9_dmft}
\end{figure*}

We present supplemental information for the Fermi surface in Fig.~\ref{Fig9_dmft}. That is, we plot it for other $k_z$ momenta than the $k_z=0$ of Fig.~\ref{Fig5_H4_FS_DMFT}.

\subsection{IV.~Lattice expansion in LaNiO$_2$H$_{\delta}$}

\begin{figure}
\includegraphics[width=0.45\textwidth]{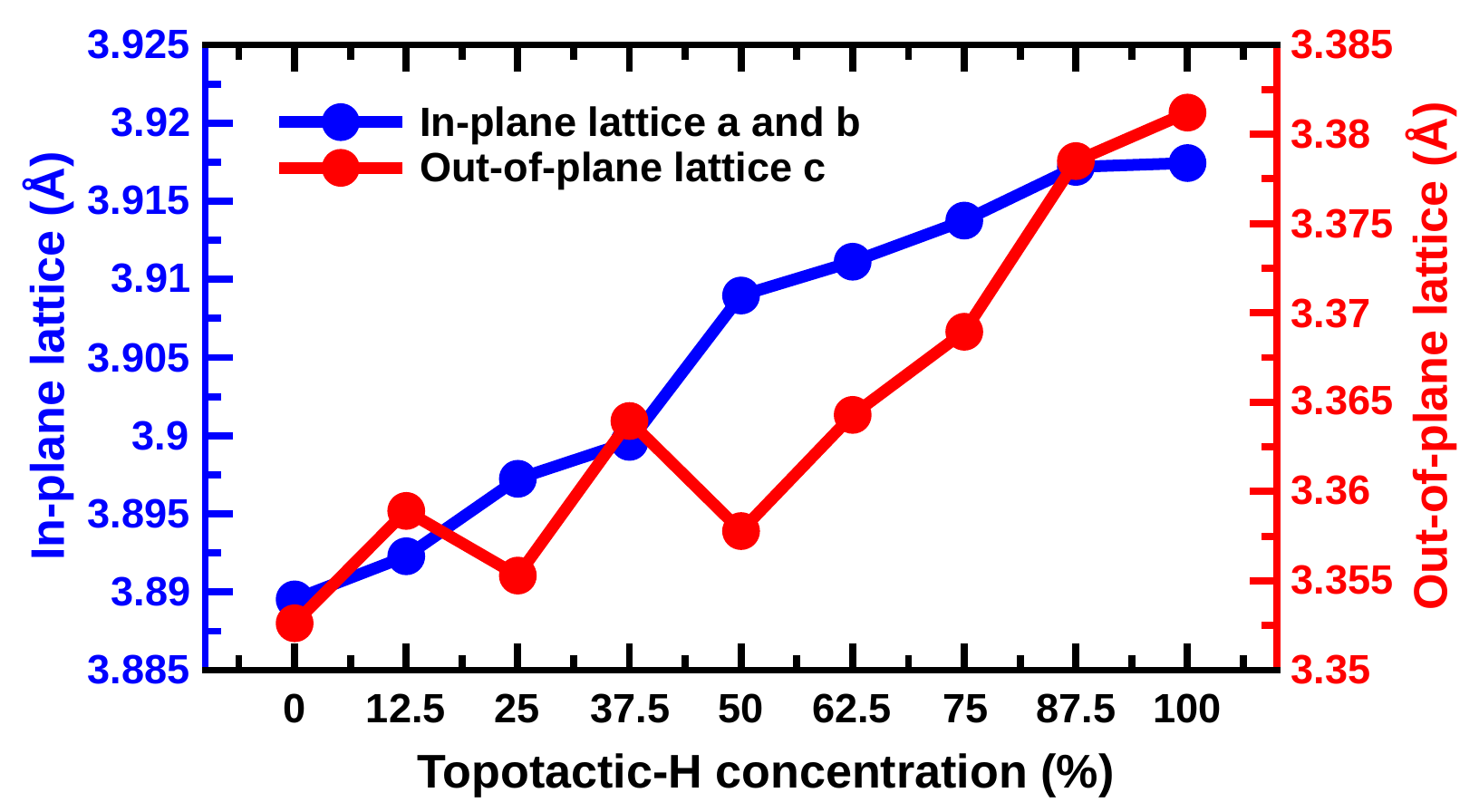}
\caption{DFT calculated lattice constants of LaNiO$_2$H$_{\delta}$ ($\delta$=0\%-100\%).}
\label{FigB5}
\end{figure}

\ls{Further,  in Fig.~\ref{FigB5} we show the in-plane and out-of-plane lattice constants, obtained with  DFT-PBESol, $a$ ($=b$) and $c$ for LaNiO$_2$H$_{\delta}$. The in-plane lattice constant $a$ (and $b$) increases from 3.889\,\AA~($\delta$=0\%) to 3.917\,\AA~($\delta$=100\%), increasing by 0.7\%. 
Let us emphasize that these are the extreme changes, for a complete saturation with topotactic-H.  For a possibly more realistic H concentration of up to 25\%, there is only a fraction of this change of lattice parameters, see Fig.~\ref{FigB5}.
And the out-of-plane lattice $c$ increases from 3.337\,\AA~to 3.381\,\AA, increasing by 1.3\%. One notable result is that both at 25\% and 50\%, there exists a reduction of the out-of-place lattice. Such a reduction is due to the formation of H-chain(s) in LaNiO$_2$H$_{0.125}$ and LaNiO$_2$H$_{0.5}$. Once the third chain forms in LaNiO$_2$H$_{0.75}$, this effect is barely visible anymore. The (quasi-)monotonously increasing in-plane and out-of-plane lattice constants demonstrate the lattice expansion induced by topotactic-H.}

\end{document}